\documentclass[conference]{IEEEtran}
\usepackage{amsmath,amsfonts}
\usepackage{algorithmic}
\usepackage{algorithm}
\usepackage{array}
\usepackage[caption=false,font=normalsize,labelfont=sf,textfont=sf]{subfig}
\usepackage{textcomp}
\usepackage{stfloats}
\usepackage{url}
\usepackage{verbatim}
\usepackage{graphicx}
\usepackage{cite}
\usepackage{comment}
\usepackage{pifont} 
\usepackage{caption}

\usepackage{multirow} 
\usepackage{booktabs}
\usepackage{caption}
\usepackage[table,xcdraw]{xcolor}
\usepackage{geometry}
\begin{document}


\title{Hiding in Plain Sight: An IoT Traffic Camouflage Framework for Enhanced Privacy}

\author{
    Daniel Adu Worae\\
    University of Notre Dame\\
    \texttt{dworae@nd.edu}
    \and
    Spyridon Mastorakis\\
    University of Notre Dame\\
    \texttt{mastorakis@nd.edu}
}


\maketitle

\begin{abstract}

The rapid growth of Internet of Things (IoT) devices has introduced significant challenges to privacy, particularly as network traffic analysis techniques evolve. While encryption protects data content, traffic attributes such as packet size and timing can reveal sensitive information about users and devices. Existing single-technique obfuscation methods, such as packet padding, often fall short in dynamic environments like smart homes due to their predictability, making them vulnerable to machine learning-based attacks. This paper introduces a multi-technique obfuscation framework designed to enhance privacy by disrupting traffic analysis. The framework leverages six techniques—Padding, Padding with XORing, Padding with Shifting, Constant Size Padding, Fragmentation, and Delay Randomization—to obscure traffic patterns effectively. Evaluations on three public datasets demonstrate significant reductions in classifier performance metrics, including accuracy, precision, recall, and F1 score. We assess the framework's robustness against adversarial tactics by retraining and fine-tuning neural network classifiers on obfuscated traffic. The results reveal a notable degradation in classifier performance, underscoring the framework's resilience against adaptive attacks. Furthermore, we evaluate communication and system performance, showing that higher obfuscation levels enhance privacy but may increase latency and communication overhead.

\end{abstract}



\section{Introduction}
The widespread adoption of IoT devices has surged in both industrial and smart home settings\cite{perera2022iot}. However, this shift has also unveiled new vulnerabilities due to various inherent design flaws in IoT devices. The well-known Mirai Distributed Denial of Service (DDoS) attack \cite{antonakakis2017understanding} has highlighted the critical need for enhanced security measures for IoT devices. As a result of this incident, studies have been initiated to investigate possible vulnerabilities that adversaries could exploit. Beyond active attacks like denial of service (DoS), adversaries may engage in passive eavesdropping to capture and analyze network traffic of IoT devices, which allows them to access sensitive information about the device or its user \cite{sivanathan2017characterizing, meidan2017profiliot}. This is made possible by the help of traffic analysis techniques. 

Traffic analysis, while essential for tasks like intrusion detection and quality of service monitoring, can also be exploited to invade user privacy \cite{zhang2018homonit}. Recent studies have shown that machine learning-based passive traffic analysis can reveal sensitive details about IoT devices and user activities, even with encrypted data \cite{wang2020fingerprinting, alshehri2020attacking}. By analyzing metadata such as packet sizes and flow patterns, these methods can accurately identify device types and usage patterns. For example, increased data transmission from a smart fitness device in the early morning may indicate a regular exercise routine, potentially exposing the homeowner's habits and compromising their privacy and security\cite{acar2020peek}.

\footnote{Upon acceptance of this paper, we will release our code as open source to support transparency and advance research in the field.}

\begin{figure}[h]
    \centering
    \includegraphics[width=3.45in]{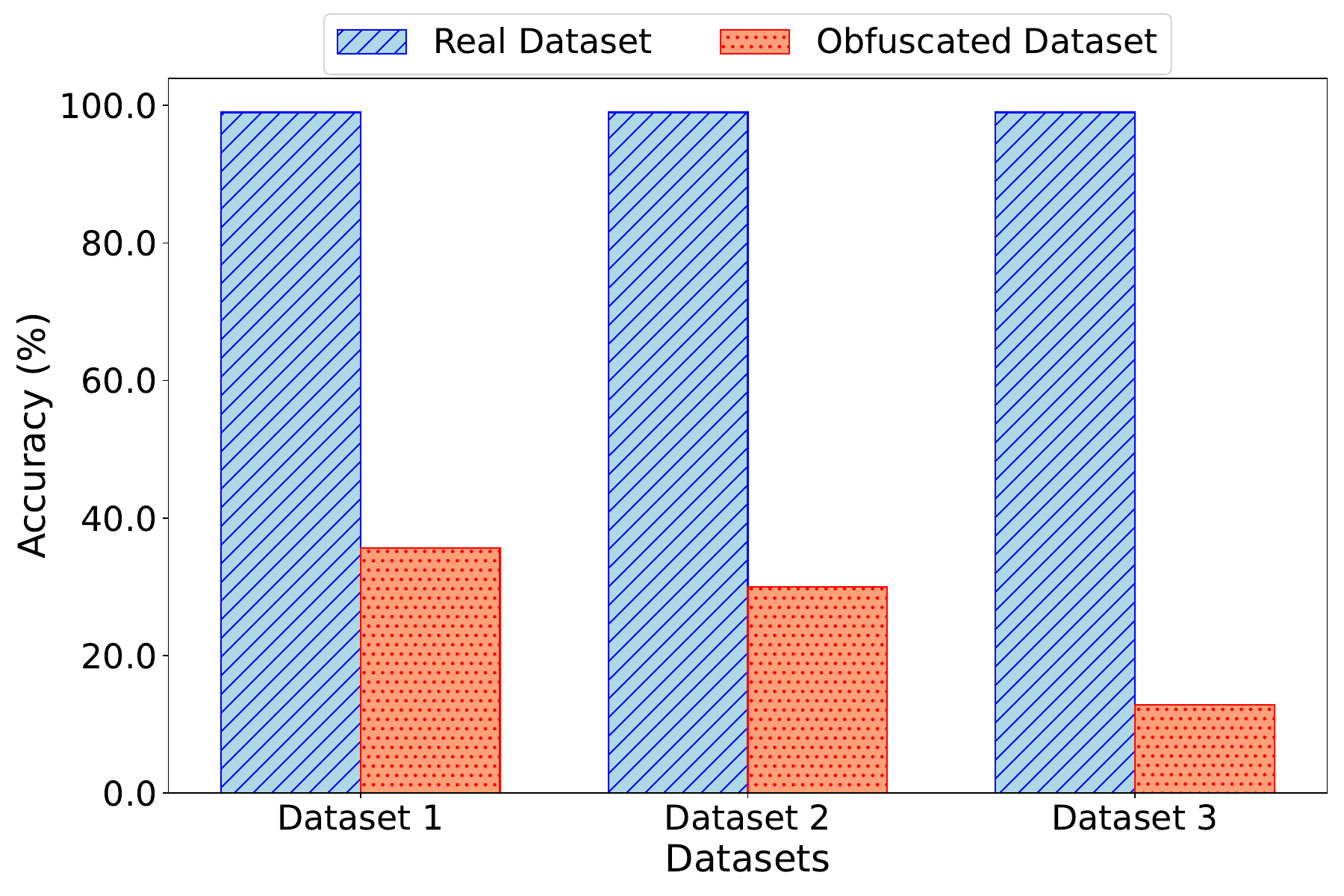}  
    \caption{Accuracy of machine learning models for IoT device classification on unaltered (real dataset) vs. obfuscated traffic. The significant drop in accuracy in classifying the obfuscated traffic highlights the effectiveness of our obfuscation framework.}
    \label{fig1}
\end{figure}

In contrast, research in adversarial machine learning explores strategies to disturb the input of trained machine learning models, which significantly lower the accuracy of classifiers. This technique is referred to as \textit{traffic obfuscation}. Traffic obfuscation is a critical area in IoT networks that addresses the noted security concerns. Obfuscation techniques disguise the characteristics of network traffic to prevent attackers from gleaning sensitive information through traffic analysis. These techniques encompass a wide range of strategies, such as encryption, traffic padding, timing and order manipulation, and false packet injection. Recent findings have indicated that such tactics effectively decrease the detection accuracy of widely utilized Machine learning-based classifiers for analyzing IoT traffic \cite{alshehri2020attacking, acar2020peek, apthorpe2018keeping, perera2022iot}. 

While these techniques have shown efficacy in enhancing the privacy and security of IoT systems, the landscape of cyber threats continues to evolve at a rapid pace. Adversaries constantly refine their methods and employ more sophisticated machine learning models to circumvent single-technique traffic obfuscation frameworks. The threat intensifies as attackers fine-tune their models to learn and adapt to traffic patterns, even when obfuscation techniques are applied. Single-technique obfuscation methods are prone to inference when attackers use powerful machine-learning models, particularly through fine-tuning or incremental training on obfuscated traffic. This necessitates a shift towards comprehensive security measures. To address these challenges, we explore the following research questions:
\begin{itemize}
    \item In the face of adversaries equipped with machine learning capabilities, to what extent can practical traffic obfuscation techniques effectively protect the privacy of IoT devices?
    \item Considering the resource-constrained nature of IoT devices, what are the trade-offs between implementing privacy safeguards and the overhead incurred by IoT devices?  
\end{itemize}
This paper introduces an advanced obfuscation framework designed to significantly enhance the privacy of IoT systems. Unlike single-technique approaches, our framework integrates multiple obfuscation methods to provide a more comprehensive defense. Diversifying traffic patterns, effectively diminishes the accuracy of machine learning-based traffic analysis, thereby improving IoT privacy. As shown in Fig. 1, machine learning models trained on unaltered IoT traffic to classify devices experience a significant drop in accuracy when our obfuscation framework is applied. The obfuscation disrupts the traffic patterns, making it much harder for the models to accurately classify the obfuscated traffic, demonstrating the framework's effectiveness in enhancing privacy and preventing accurate device identification. 

\subsection{Contribution}
This work offers substantial contributions to IoT privacy enhancement. The principal contributions are summarized as follows:

\begin{itemize}
    \item We introduce a novel framework that integrates multiple obfuscation techniques, including Padding, Padding and XORing, Padding and Shifting, Constant Size Padding, Fragmentation, and Delay Randomization. This framework effectively obfuscates traffic patterns to protect against traffic analysis attacks.
    \item We rigorously evaluate the framework using three public IoT traffic datasets. This evaluation not only assesses its impact on classifier metrics such as accuracy, precision, recall, and F1 score but also examines the framework's effects on communication performance, including data overhead, and system performance metrics like execution time, memory usage, and CPU load.
    \item We assess the robustness of our framework through incremental training and fine-tuning on neural network model with obfuscated traffic. This evaluation determines the extent to which our obfuscation techniques degrade classifier performance, even when models are retrained and fine-tuned to the obfuscated data, thereby demonstrating resilience against adversarial tactics.
\end{itemize}

The rest of our paper is organized as follows: In Section II, we present a literature review of the study. In Section III, we present the threat model of our framework. In Section IV, we describe the experimental system flow of our obfuscation framework. Section V presents experimental results and an analysis of our proposed framework. In Section VI, we present the numerical results of our proposed framework. Finally, we conclude in Section VII. 

\section{Literature Review}
\label{sec: Lit}
The integration of IoT systems into daily life marks a significant digital advancement, connecting devices for efficient data exchange but raising serious privacy concerns, such as identity theft and physical threats. Traffic obfuscation has emerged as a key strategy to protect privacy by concealing communication patterns in IoT ecosystems. This review explores key techniques in traffic obfuscation, including (\ref{sec: encryption}) IoT traffic obfuscation through encryption, (\ref{sec: padding}) traffic padding, (\ref{sec: Timing_Order}) timing and order obfuscation, and (\ref{sec:Injection}) synthetic packet injection for obscuring IoT traffic.

\subsection{IoT traffic Obfuscation through Encryption}
\label{sec: encryption}
In the realm of Internet of Things (IoT) security, the quest for a balance between privacy preservation and effective network traffic inspection presents a formidable challenge. 
IoT traffic obfuscation through encryption refers to the process of disguising or hiding the actual data being transmitted by Internet of Things (IoT) devices by using encryption techniques. The main goal is to protect the privacy and security of the data as it travels across networks, which makes it difficult for unauthorized parties to interpret or make use of the information even if they manage to intercept it.   
   
There have been multifaceted approaches to encryption-based methods for obfuscating IoT traffic. Each of the methods designs innovative solutions to safeguard privacy and at the same time, maintains the functionality of network security measures \cite{9723011, 7845477, REN2021105, 10.1145/2785956.2787502, 194934}. These methods collectively underscore the potential that comes from using encryption and novel protocols to enable deep packet inspection (DPI) of encrypted traffic. They highlight the efficacy of such methods in preserving privacy and security, and present systems that can perform DPI without the need for decryption. This method ensures network security and maintains traffic confidentiality.
    
However, insights from recent studies present a critical vulnerability in encryption-based privacy methods. They demonstrate that encryption alone is not enough to prevent privacy breaches. Certain traffic characteristics such as packet headers and inter-arrival times remain exposed despite encryption \cite{msadek2019iot, apthorpe2017spying, apthorpe2017smart, 10.1145/3559613.3563191, apthorpe2018keeping}. These can be exploited through machine learning and traffic analysis to infer user activities. This is particularly concerning for IoT environments, where privacy breaches can impact the personal realm of smart homes and devices.

The contrast between the optimistic view of encryption-based methods and cautionary evidence from later studies underscores a critical insight: achieving robust privacy in IoT networks requires more than just encryption. It demands a holistic approach that combines encryption with other mechanisms to obfuscate or eliminate sensitive information leakage through traffic metadata.

\subsection{Traffic Padding} 
\label{sec: padding}
The objective of traffic padding is to assess how the incorporation of dummy data to packets obscures real user activities and data patterns, evaluating their effectiveness in counteracting traffic analysis attacks. Traffic padding involves adding extra bytes to packet data to obfuscate its true size. This complicates traffic analysis attempts that could otherwise reveal sensitive user information. In recent years, the development of traffic padding techniques has become central which enhances privacy in Internet of Things (IoT) networks. Various studies have introduced innovative approaches to traffic padding to mitigate the effectiveness of traffic analysis attacks\cite{8538744, datta2018developer, xiong2018defending, 9825227, 9750450, 6234422}. 

For instance, adaptive packet padding methods dynamically adjust the amount of padding based on real-time network conditions, which balances privacy preservation with network performance efficiency\cite{9203848}. Meanwhile, lightweight padding mechanisms introduce minimal overhead solutions that obscure packet sizes\cite{8538744}. 

Other analyses highlight the limitations of traffic padding techniques in IoT networks, showing that sophisticated methods can still identify patterns in encrypted traffic and reveal user activities and device types\cite{10.1145/3559613.3563191}. To address privacy concerns, Dynamic STP (DSTP) was introduced, incurring significantly less per-packet overhead. Existing traffic obfuscation methods primarily alter traffic from IoT devices, neglecting incoming server traffic, which allows successful inference of IoT activities. Effective obfuscation must alter network traffic in both directions between IoT devices and servers.

\subsection{Timing and Order Obfuscation} 
\label{sec: Timing_Order}

Traffic shaping involves manipulating the timing and sequencing of network traffic to prevent unauthorized inference of user activities from data patterns. This technique aims to disguise the actual usage patterns of internet-connected devices by altering the observable characteristics of traffic flows, thereby enhancing privacy.

Time and order obfuscation techniques stand out for their ability to mitigate the risks associated with traffic analysis attacks. By introducing variability in packet timing and ordering, these methods effectively blur the distinctive traffic patterns that could otherwise be used to infer sensitive user information. This is particularly crucial in smart home environments, where the nature and timing of device communications can reveal intimate details about the inhabitants' daily routines and preferences.

Traffic shaping provides innovative ways of effectively preserving IoT privacy through timing and order obfuscation of IoT traffic\cite{5961736, apthorpe2017spying, 10.1145/3374664.3375723, hussain2021dark, 9322070, apthorpe2018keeping}. The obfuscation of time and order in IoT traffic has emerged as a promising strategy that protects against privacy vulnerabilities inherent in the encrypted traffic of IoT devices. By obfuscating the metadata, such as transmission timings, that can be exploited for traffic analysis, time and order obfuscation serves as a critical tool in the privacy-preserving toolkit for IoT privacy and security. 

\subsection{Injection of Synthetic packets} 
\label{sec:Injection}

In the realm of IoT privacy protection, the injection of synthetic packets has emerged as a pivotal strategy. This method generates and sends dummy packets that blend seamlessly with legitimate traffic and mask the true nature of network activities\cite{8730787, zhu2021smart, 9609087, apthorpe2017closing}. By imitating traffic patterns of actual devices, synthetic packet injection complicates the task of identifying device states and user behaviors, which presents a significant hurdle for unauthorized data analysis.

The effectiveness of synthetic packet injection lies in its ability to create a layer of ambiguity over network communications. This not only prevents the accurate identification of devices but also shields user activities from being monitored or inferred\cite{8704324, 9464026, zhang2023novel}. The strategic insertion of these dummy packets into the network traffic serves as a proactive measure against privacy breaches. This ensures that the data being transmitted does not inadvertently reveal sensitive information about the users or their habits. The key challenge lies in accurately timing the introduction of decoys to ensure they are indistinguishable from real events, which maintains parallelism with actual activities without logical inconsistencies.

While these methods have advanced IoT privacy and security, several key gaps remain. First, many solutions rely on single-technique obfuscation methods, such as padding, which are often predictable. Sophisticated adversaries can fine-tune machine learning models to bypass these defenses. Our multi-technique framework addresses this by disrupting multiple traffic dimensions making it significantly harder for adversaries to adapt. Moreover, most approaches focus on specific traffic features like packet size or timing, leaving other metadata exposed. Our framework uses diverse techniques to target multiple aspects of traffic, offering broader protection against traffic analysis.

Lastly, current methods lack rigorous testing against adaptive adversaries, who can retrain models on obfuscated data. Our framework is tested against such adversaries and demonstrates resilience by significantly degrading classifier performance even after adversarial retraining.

In light of these gaps, our work advances the state of the art by providing a comprehensive, multi-technique traffic obfuscation framework that not only mitigates the limitations of single-technique methods but also demonstrates robustness against adaptive adversarial strategies.

\section{Threat model}

This paper addresses the privacy risks associated with network traffic from smart home devices, which can inadvertently reveal sensitive information about both the devices and their users. Despite the widespread use of encryption in smart home communications, adversaries can still glean valuable insights from traffic metadata, such as packet lengths and the number of packets per flow. We consider two types of adversaries: WAN sniffers and Wi-Fi sniffers.

A WAN sniffer monitors traffic between the home router and the ISP network. In a typical home network, Network Address Translation (NAT) manages multiple devices sharing a single public IP address. The router replaces the device's private IP address and MAC address with its own public IP and MAC address, masking the original device's identity from external networks. As a result, WAN sniffers can only see the public IP address of the router and the IP headers of packets, lacking access to the original MAC addresses of devices behind the router\cite{trimananda2019pingpong}. They cannot directly identify individual devices within the home network but can observe traffic patterns and volumes associated with the home network's public IP. While this limits their ability to identify specific devices, they can infer overall activity levels and possibly deduce when the network is in use.

A Wi-Fi sniffer intercepts encrypted IEEE 802.11 traffic within the home network. Although Wi-Fi traffic is encrypted with protocols like WPA2, certain metadata remains unencrypted, including MAC addresses, packet sizes, and timing information. Wi-Fi sniffers cannot access the content of the communications but can analyze this metadata. Unlike WAN sniffers, Wi-Fi sniffers within the home network can see the MAC addresses of all devices communicating over Wi-Fi, allowing them to distinguish between different devices based on their unique MAC addresses. By examining the timing and size of packets, Wi-Fi sniffers can infer the type of activity occurring on each device. This could reveal details such as when users are typically home, their daily routines, and the types of devices they use.

For both types of adversaries, we assume they have prior knowledge of the specific smart home devices they aim to monitor. They can train detection systems on similar devices offline, capturing unique traffic signatures. These signatures can then identify targeted devices in real-time by analyzing consistent metadata patterns. This threat model highlights the privacy vulnerabilities in smart home networks and underscores the need for obfuscation techniques capable of altering traffic patterns to thwart traffic analysis methods, thereby protecting user privacy.

\section{Experimental System Flow of the Proposed Obfuscation Framework}

\subsection{System Design}
In this study, we propose a novel approach to enhancing IoT privacy through comprehensive traffic obfuscation techniques as shown in Figure 2. We aim to prevent adversaries from inferring device types and user activities in smart home environments by reducing machine learning-based traffic analysis evaluation metrics. This system design encompasses several components and their interactions, forming a cohesive structure that ensures data privacy. Figure 3 shows the system design.

\subsubsection{IoT Devices and Traffic Generation} The framework supports a diverse range of IoT devices, from consumer electronics like smart home gadgets to industrial sensors. These devices generate network traffic, including data packets (which carry the payload, such as sensor readings) and control packets (which manage device communication and configurations).

\begin{figure}[h]
    \centering
    \includegraphics[width=3.45in]{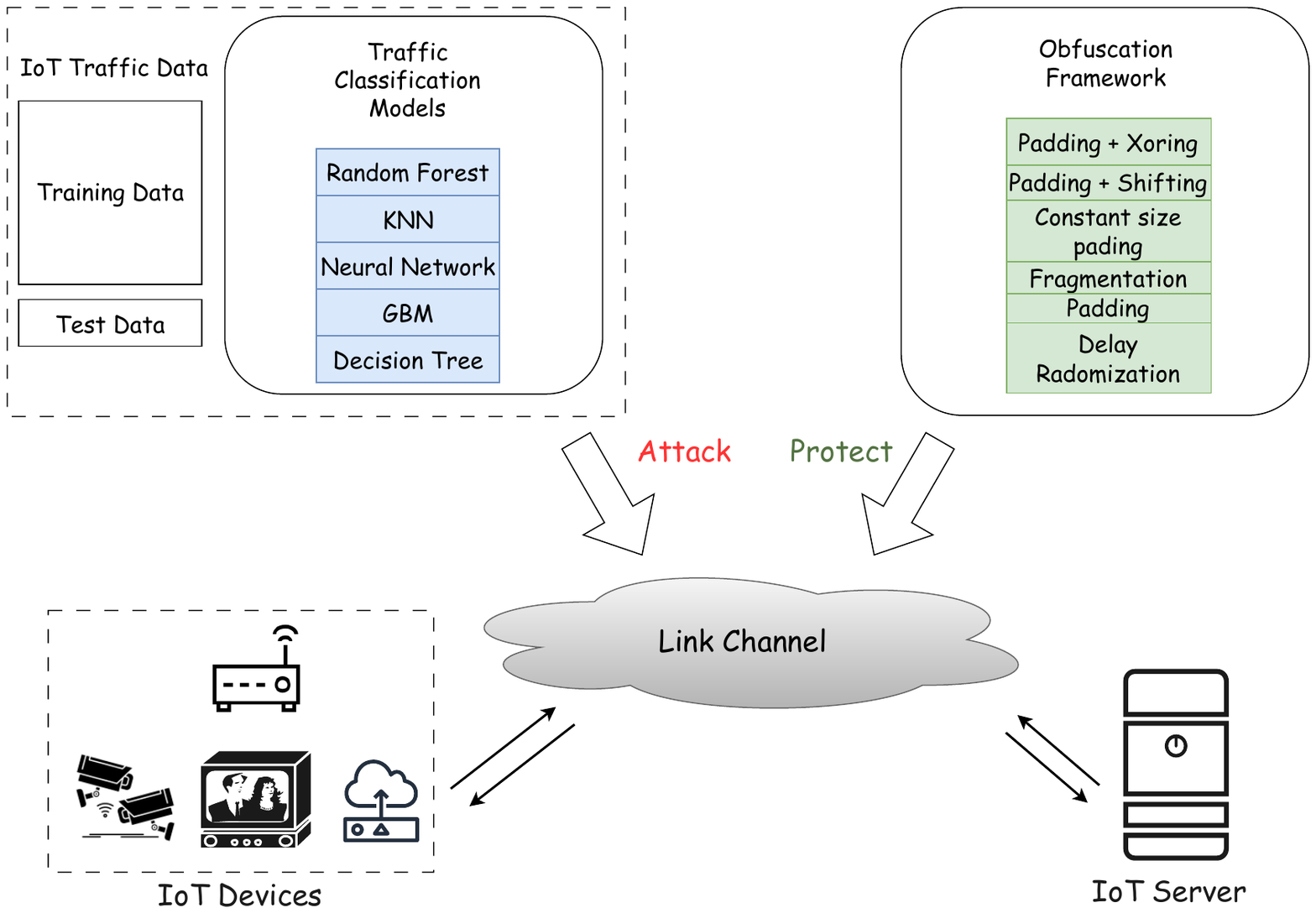} 
    \caption{Obfuscation Techniques against Traffic Classification Models}
    \label{fig2}
\end{figure}

\begin{figure*}[h]
    \centering
    \includegraphics[width=4.8in]{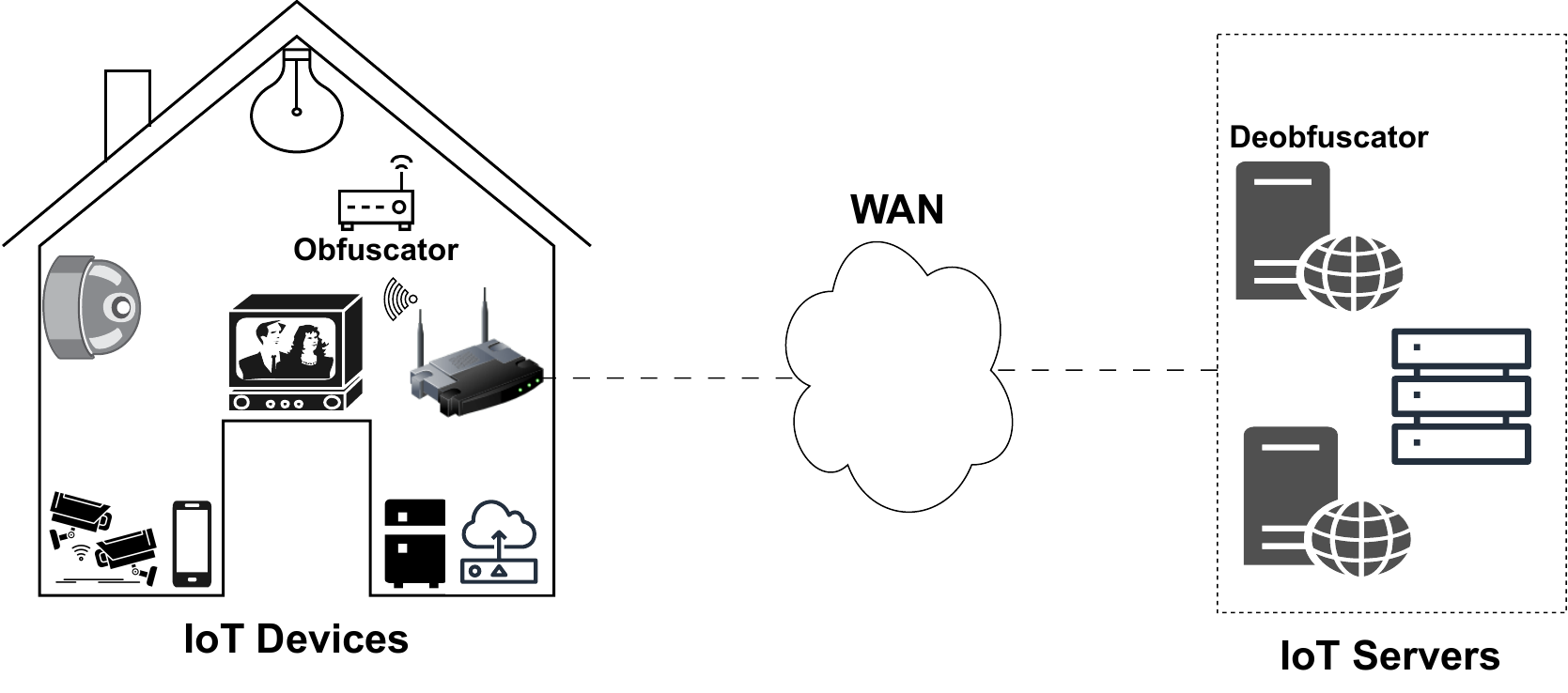} 
    \caption{System Design}
    \label{fig3}
\end{figure*}

\subsubsection{Obfuscation Framework} The core of the system is an obfuscation framework implemented on a Raspberry Pi, which connects to the network router via its Wi-Fi interface. This implementation leverages Python for its ease of use and compatibility with a wide range of IoT devices and applications. The Raspberry Pi intercepts and processes all outgoing traffic from the IoT devices, which, in this case, involves replayed traffic, applying a suite of obfuscation techniques designed to mask specific aspects of the traffic. The framework incorporates six obfuscation techniques, which are applied to the traffic to protect against potential threats, even in the presence of adaptive adversaries retraining their models with obfuscated traffic. We strategically combine some of the techniques to disrupt multiple traffic characteristics, making it harder for adversaries to adapt, even with fine-tuned models.  The six obfuscation techniques used are detailed below:

\begin{itemize}
    \item \textbf{Padding:} This technique involves adding random bytes to the payload of each packet, increasing its size. By introducing arbitrary data, we obscure the original packet size, making it difficult for adversaries to deduce meaningful information from traffic patterns. 
    \item \textbf{Padding and XORing:} This method pads the payload with random bytes and generates a second set of random bytes equal in length. The original payload is then XORed with the second set, creating a dual-layer obfuscation that both increases the payload size and conceals its content. This approach provides strong protection against payload inspection and analysis, ensuring that even in unencrypted packets, the XOR operation effectively hides the data from adversaries.
    \item \textbf{Padding and Shifting:} This technique combines padding with random shifting of the padded payload. After adding random bytes, the entire payload is shifted in a random order, disrupting the natural data structure. This adds complexity to traffic patterns, making it difficult for adversaries to conduct accurate analysis. In the absence of encryption, the random shifting further obscures the content, providing strong protection against unauthorized inspection and analysis.
    \item \textbf{Constant Size Padding:} This method pads all packets to a uniform size based on the largest packet in the traffic. For example, if the largest packet is 100 bytes, all packets are padded to match this size. This uniformity eliminates size-based inference attacks, as adversaries cannot distinguish between different types of traffic based on packet size.
    \item \textbf{Fragmentation:} This technique divides the payload into multiple smaller fragments of varying sizes. Breaking the payload into irregular pieces prevents adversaries from correlating packet sizes with specific activities or device types. This creates a fragmented traffic flow, making it difficult to analyze and interpret.
    \item \textbf{Delay Randomization:} This technique adds random delays to packet transmission, disrupting temporal patterns that adversaries could use to infer activity or identify devices. Introducing unpredictability in timing obscures traffic patterns and enhances privacy.
\end{itemize}

\subsubsection{Deobfuscation} 
Each obfuscated packet is equipped with a recovery header, which adds a modest overhead of 2 to 4 bytes. This recovery header is crucial for guiding the accurate reversal of the obfuscation technique applied. The header contains an Obfuscation Technique ID, allowing the de-obfuscation framework to identify which specific technique was used on the packet. Additionally, the header includes metadata specific to the obfuscation method, such as padding length for padding techniques, XOR keys for XOR-based obfuscation, shift amounts for shifting-based methods, or fragment sequence numbers for fragmentation.

This metadata is essential for enabling the framework to reverse the obfuscation with precision, ensuring accurate recovery of the original packet. Once a packet is received, the de-obfuscation process begins by extracting the recovery header. This header serves as a blueprint that informs the framework of the sequence of transformations applied to the packet. The framework identifies the obfuscation technique via the Obfuscation Technique ID and retrieves any parameters needed for the reversal, such as padding length or shift amounts. Using this information, the framework methodically reverses each transformation in the correct sequence. For instance, padding is removed by referencing the exact length specified in the metadata, XOR operations are reversed using the provided key and shifted data is re-aligned to its original sequence based on the shift parameters detailed in the recovery header.

To improve security and avoid predictable header locations, the recovery header's position is dynamically determined by a predetermined algorithm using packet metadata, like sequence numbers or size. This approach enhances security by making the header's location harder to predict while ensuring synchronization between sender and receiver for seamless de-obfuscation without complex encryption.

Once de-obfuscation is complete, the IoT device processes the restored packet by inspecting its payload, executing any relevant actions or commands, or forwarding the data as needed for further processing or delivery.



\subsubsection{Packet Transmission and Reception} The obfuscated packet is transmitted over the network, with the modifications ensuring that any intercepted traffic remains resistant to analysis. At the receiving end, authorized servers or devices use the metadata to discard cover traffic and recover the original messages, ensuring that only intended recipients can access the unaltered data. The system relies on a shared understanding between communicating parties about the obfuscation parameters, which are periodically updated to maintain security and prevent predictability.

\subsection{Data Preparation and Model Assessment}
In this phase, we capture and preprocess both unobfuscated and obfuscated IoT traffic data to ensure consistency and prepare the data for training and evaluating machine learning models.

\subsubsection{Data Preprocessing} We capture IoT traffic to analyze data patterns generated by various devices. The principle of "garbage in, garbage out" underscores the importance of data preprocessing in any machine learning task\cite{danso2023transferability}. Training a model with unclean data results in poor model performance and irrelevant analysis. In this stage, we address all missing data records for the measured variables. We apply Z-score normalization, also known as standardization, to scale our data into a regularized range. Z-score normalization adjusts features according to the standard normal distribution with a mean ($\mu$) of 0 and a standard deviation ($\sigma$) of 1, calculated using
\begin{equation}
z = \frac{x_i - \mu}{\sigma}
\end{equation}


\subsection{Feature Selection}
Feature selection aims to identify a subset of input variables that effectively represent the entire dataset, minimizing noise and irrelevant attributes while maintaining accurate prediction results. Additionally, feature selection reduces computational demands, mitigates the "curse of dimensionality," and improves generalization performance. In this study, we employed the analysis of variance (ANOVA) feature selection method. ANOVA is a statistical technique used to evaluate one or more dependent variables under different conditions defined by one or more measurements. This method assumes that the variables follow a Gaussian distribution and that there is a linear relationship between each feature and the target variable.

Within ANOVA, the F-test is employed to assess whether significant differences exist between the groups. If there is no substantial difference between the groups and all variances are equal, the F-ratio will be close to 1. This indicates that the attribute has no impact on the response and is not useful for model training. The SelectKBest module utilizes the score values generated by the ANOVA module to select the optimal set of features for training.

\subsection{Model Training}
In this phase, we train five machine learning models to profile and identify different device types: Decision Tree (DT), Random Forest (RF), Gradient Boosting Machine (GBM), K-Nearest Neighbor (kNN), and Neural Network (NN). The hyperparameters utilized in the classifiers are outlined in Table I. 

We use these models to establish a baseline for performance metrics using normal IoT traffic. To further test the efficiency of our obfuscation techniques, we perform fine-tuning and incremental training specifically on the Neural Network. This additional step helps us evaluate the effectiveness of our obfuscation methods against more sophisticated machine learning models and ensures that our approach effectively reduces the ability of these models to accurately analyze and infer sensitive information from the obfuscated traffic.

\begin{table}[ht]
\centering
\caption{Model Hyperparameters}
\label{table:model-parameters}
\begin{tabular}{
    |>{\raggedright}p{3.4cm}|
    >{\raggedright\arraybackslash}p{4.5cm}|}
\hline
\rowcolor{lightgray}
\textbf{Model} & \textbf{Parameters} \\
\hline
Random Forest & n\_estimators=53, random\_state=30 \\
\hline
K-Nearest Neighbors & n\_neighbors=5, algorithm='auto', metric='minkowski', weights='uniform' \\
\hline
Decision Tree & criterion='gini', splitter='best', max\_depth=None, random\_state=30, min\_samples\_split=2 \\
\hline
Gradient Boosting Machine (GBM) & learning\_rate=0.1, n\_estimators=100, subsample=1.0, max\_depth=3, random\_state=30 \\
\hline
Neural Network (NN) & 3 hidden layers (64 neurons each, ReLU activation), Adam optimizer, categorical cross-entropy loss, 330 epochs, batch size=32 \\
\hline
\end{tabular}
\captionsetup{justification=centering}
\caption*{Summary of model hyperparameters used in the experiments.}
\end{table}

\subsection{Testing and Evaluation} In this module, we assess the effectiveness of our obfuscation techniques by evaluating the performance of the trained models on preprocessed obfuscated traffic. Our evaluation focuses on several key metrics: Accuracy, Precision, Recall, and F1 Score. These metrics provide an understanding of how well the models perform in the presence of obfuscation and the robustness of our techniques in preventing traffic analysis.

We begin by establishing baseline performance metrics for the machine learning models using normal IoT traffic. This initial evaluation provides a reference point for assessing the impact of the obfuscation techniques. After applying the six obfuscation techniques to the IoT traffic, we test the trained models against the obfuscated traffic to measure their performance. The performance metrics are compared to the baseline to quantify the effectiveness of the obfuscation techniques in reducing the models' ability to accurately analyze the traffic.

Additionally, to further test the robustness of our obfuscation methods, we perform fine-tuning and incremental training on the neural network using the obfuscated traffic. This step evaluates how well the neural network adapts to the obfuscated data and how the obfuscation techniques hold up against a model that undergoes continuous improvement.

\section{Experimental Results and Analysis}
To evaluate our comprehensive obfuscation framework, we used packet capture (pcap) files from three distinct datasets—IoT-AD\cite{zahan2023iot}, IoT Sentinel\cite{miettinen2017iot}, and UNSW\cite{sivanathan2018classifying}. These datasets provide a diverse range of IoT traffic data, ensuring comprehensive testing across different environments. Each dataset was used separately to test our framework, allowing for a thorough assessment of its effectiveness across various scenarios.

The pcaps were replayed using a laptop on the same local network as a Raspberry Pi. A Python program, utilizing Scapy, was used on the laptop to replay the packets. The Raspberry Pi, with a 2.4GHz 64-bit quad-core CPU, 8GB of RAM, and running the 64-bit Raspberry Pi OS, captured the replayed traffic and applied the obfuscation techniques. The framework was implemented in Python, leveraging the capabilities of the Scapy library for packet manipulation and transmission.

Scapy is an interactive packet manipulation program that enables users to send, sniff, dissect, and forge network packets. This functionality was critical for our prototype, allowing us to dynamically alter packet attributes and forge new packets to obscure traffic patterns effectively.

The Raspberry Pi captured the replayed traffic, which included various IoT device communications. Our obfuscation framework was then applied to the captured traffic, using Scapy to modify packet attributes such as payload size and packet timing. After applying the obfuscation techniques, the obfuscated packets were sent to another PC, acting as a server, where deobfuscation processes were applied. This step was crucial for evaluating the full cycle of our framework, from obfuscation to deobfuscation, ensuring that the data could be restored while maintaining privacy.

The primary objective of this evaluation is to determine whether the obfuscation framework can effectively reduce the evaluation metrics of IoT traffic classifiers. The framework should be able to obfuscate packets with minimal impact on network performance metrics. We conducted two sets of experiments: one to assess the performance of the traffic classifiers with and without the obfuscation mechanisms, and another to measure the communication overhead. In both scenarios, the obfuscation framework was deployed on the Raspberry Pi, while the laptop handled the packet replay.

The effectiveness of our obfuscation framework is assessed using several metrics: accuracy, precision (the percentage of correctly classified positive instances out of all instances classified as positive), recall (the percentage of correctly classified traffic for each device), and F1 score (the harmonic mean of precision and recall). Additionally, the impact on communication performance is measured through the bytes added, execution time, and delay. We also evaluate system performance by monitoring CPU and memory usage during the application of the obfuscation techniques. The following experiments outline these assessments.

\subsection{Privacy-Preserving Experiments}
We assume that a network observer has access to datasets containing various IoT packet features, such as packet length, flow length, flow time, and destination port, and uses this data to train machine learning classification algorithms. By employing these classifiers, the observer can identify IoT devices and infer user activities within the network of their targets. This type of attack requires minimal effort and low-level computational resources, such as a personal computer.

In our setup, a Python program utilizing Scapy replayed the IoT traces from the client to the server. The obfuscation framework was deployed on a Raspberry Pi. The Raspberry Pi captured the traffic and applied the obfuscation framework to the traffic. The obfuscated traffic was captured on its Wi-Fi interface using tcpdump and saved as pcap files. This configuration effectively simulates a scenario where an observer captures all traffic originating from the victim's network, allowing for a comprehensive assessment of the obfuscation framework's effectiveness in protecting against traffic analysis attacks.

All pcap files, containing obfuscated or unobfuscated traffic, were converted into CSV format. The CSV files included various features such as delta time, destination port, packets per second, packet length, total packets per flow, conversation length, total packet length, TCP/UDP segment length, TCP/UDP stream time, and flow time of the IoT traffic. The UNSW dataset contributed 164,194 samples along with 164,194 corresponding obfuscated samples; the IoT-AD dataset provided 7,231 samples and an equal number of obfuscated samples; and the IoT Sentinel dataset included 17,003 samples, each matched with an obfuscated counterpart. The CSV files were used to train and test five widely used traffic classification classifiers: K-NN, RF, DT, NN, and GBM. These classifiers were implemented using the scikit-learn library. A stratified 10-fold cross-validation approach was employed to assess the performance of these algorithms in classifying IoT traffic, ensuring a balanced representation of each class across all folds for more reliable evaluation.

For each device in the CSV files, we used the features stated previously to identify the device and its activities from the obfuscated traffic. This allowed us to assess the performance of the obfuscation framework in concealing patterns in IoT traffic. In this scenario, we assumed that the observer was unaware of the specifics of the obfuscation framework.

We also considered a scenario where the observer has access to our obfuscation framework and can use it to obfuscate the training data. In this context, we conducted incremental training and fine-tuning on the models using obfuscated traffic to evaluate the framework's effectiveness. This experiment aimed to assess the effectiveness of the obfuscation techniques in safeguarding device types and activities, even when the observer has prior knowledge of the obfuscation framework.

\subsection{Communication and System Performance Experiments}
To evaluate the impact of our comprehensive obfuscation framework on communication and system performance, we conducted a series of experiments. The primary focus was on measuring the communication cost associated with the obfuscation techniques and the performance impact on the Raspberry Pi.
The communication performance was assessed by examining the additional bytes added to the traffic and the delay introduced by the obfuscation process, both measured as averages per packet. We quantified the average number of padding bytes added to each packet and measured the latency as the average time to apply the obfuscation techniques per packet.

System performance was evaluated by monitoring CPU usage, memory usage, and execution time, all calculated as average consumption per packet during the obfuscation process. CPU utilization and memory consumption were measured before and after applying the obfuscation techniques to determine the processing overhead introduced by the framework. It should be noted that the reported metrics reflect one-way consumption during the obfuscation process. However, these metrics are expected to be the same for de-obfuscation as the same operations (in reverse) are performed to recover the original data.
The average time to apply the obfuscation techniques per packet was recorded to assess the framework’s processing efficiency.

\begin{table*}[ht]
\centering
\caption{Performance Metrics for Various Obfuscation Techniques Across Three Datasets (IoT-AD, IoT Sentinel, UNSW)}
\label{table:performance-metrics_1}

\resizebox{1\textwidth}{!}{%
\begin{tabular}{
    |>{\raggedright}p{2cm}|
    >{\raggedright}p{2.7cm}|
    >{\centering\arraybackslash}p{2cm}|
    >{\centering\arraybackslash}p{2cm}|
    >{\centering\arraybackslash}p{2cm}|
    >{\centering\arraybackslash}p{2cm}|
}
\hline
\rowcolor{lightgray}
\textbf{Classifier} & \textbf{Obfuscation Technique} & \textbf{Accuracy (\%) (d1/d2/d3)} & \textbf{Precision (\%) (d1/d2/d3)} & \textbf{Recall (\%) (d1/d2/d3)} & \textbf{F1 Score (\%) (d1/d2/d3)} \\
\hline
\multirow{10}{*}{Neural Network} & None (Original Data) & 94.00/91.00/98.00 & 94.00/91.00/98.00 & 94.00/91.00/98.00 & 94.00/91.00/98.00 \\
& Padding & 30.00/25.00/5.00 & 6.00/22.00/17.00 & 20.00/25.00/5.00 & 9.00/12.00/1.00 \\
& Padding + XORing & 29.00/25.00/5.00 & 6.00/35.00/8.00 & 20.00/25.00/5.00 & 9.00/13.00/1.00 \\
& Padding + Shifting & 29.00/25.00/5.00 & 6.00/41.00/8.00 & 20.00/25.00/5.00 & 9.00/13.00/1.00 \\
& Fragmentation & 36.00/20.00/5.00 & 27.00/22.00/1.00 & 21.00/20.00/5.00 & 12.00/9.00/1.00 \\
& Constant Size Padding & 32.00/23.00/16.00 & 26.00/6.00/31.00 & 22.00/23.00/16.00 & 13.00/9.00/15.00 \\
& Delay Randomization & 58.00/68.00/41.00 & 68.00/71.00/50.00 & 64.00/68.00/41.00 & 63.00/66.00/38.00 \\
& Incr. Training (Avg) & 33.17/43.33/39.83 & 28.83/48.50/32.50 & 36.33/44.33/39.83 & 26.67/39.67/29.83 \\
& Fine-Tuning (Avg) & 31.67/48.50/50.33 & 32.83/50.17/49.33 & 35.50/48.50/50.33 & 28.17/43.83/45.17 \\
\hline
\multirow{6}{*}{Random Forest} & None (Original Data) & 99.59/99.79/99.97 & 99.59/99.79/99.97 & 99.59/99.79/99.97 & 99.59/99.79/99.97 \\
& Padding & 66.00/49.00/42.00 & 46.00/44.00/51.00 & 51.00/49.00/42.00 & 47.00/44.00/34.00 \\
& Padding + XORing & 65.00/47.00/42.00 & 46.00/43.00/53.00 & 51.00/47.00/42.00 & 47.00/43.00/35.00 \\
& Padding + Shifting & 65.00/47.00/41.00 & 46.00/43.00/52.00 & 51.00/47.00/41.00 & 47.00/43.00/34.00 \\
& Fragmentation & 69.00/52.00/49.00 & 47.00/54.00/55.00 & 52.00/52.00/49.00 & 48.00/47.00/43.00 \\
& Constant Size Padding & 64.00/51.00/45.00 & 39.00/49.00/57.00 & 51.00/51.00/45.00 & 44.00/42.00/39.00 \\
& Delay Randomization & 81.00/76.00/93.00 & 86.00/83.00/91.00 & 83.00/76.00/93.00 & 82.00/76.00/92.00 \\
\hline
\multirow{6}{*}{K-NN} & None (Original Data) & 92.00/91.00/97.00 & 93.00/91.00/97.00 & 92.00/91.00/97.00 & 92.00/91.00/97.00 \\
& Padding & 30.00/23.00/5.00 & 6.00/5.00/17.00 & 20.00/23.00/17.00 & 9.00/8.00/2.00 \\
& Padding + XORing & 29.00/25.00/5.00 & 6.00/42.00/10.00 & 20.00/25.00/5.00 & 9.00/13.00/1.00 \\
& Padding + Shifting & 29.00/24.00/5.00 & 6.00/16.00/1.00 & 20.00/24.00/17.00 & 9.00/11.00/2.00 \\
& Fragmentation & 36.00/19.00/5.00 & 27.00/3.00/8.00 & 21.00/19.00/5.00 & 12.00/6.00/1.00 \\
& Constant Size Padding & 32.00/23.00/8.00 & 23.00/6.00/23.00 & 21.00/23.00/8.00 & 12.00/9.00/6.00 \\
& Delay Randomization & 75.00/66.00/76.00 & 73.00/68.00/72.00 & 74.00/66.00/73.00 & 73.00/65.00/72.00 \\
\hline
\multirow{6}{*}{Decision Tree} & None (Original Data) & 99.52/99.53/99.89 & 99.52/99.53/99.89 & 99.52/99.53/99.89 & 99.52/99.53/99.89 \\
& Padding & 66.00/56.00/35.00 & 46.00/56.00/32.00 & 52.00/56.00/35.00 & 47.00/54.00/26.00 \\
& Padding + XORing & 65.00/56.00/35.00 & 45.00/56.00/31.00 & 52.00/56.00/35.00 & 47.00/54.00/25.00 \\
& Padding + Shifting & 65.00/54.00/35.00 & 45.00/55.00/32.00 & 52.00/54.00/35.00 & 47.00/53.00/26.00 \\
& Fragmentation & 68.00/57.00/44.00 & 45.00/59.00/46.00 & 51.00/57.00/46.00 & 47.00/55.00/38.00 \\
& Constant Size Padding & 65.00/39.00/32.00 & 39.00/36.00/28.00 & 52.00/39.00/32.00 & 44.00/35.00/23.00 \\
& Delay Randomization & 68.00/79.00/82.00 & 61.00/83.00/81.00 & 60.00/79.00/82.00 & 59.00/79.00/81.00 \\
\hline
\multirow{6}{*}{Gradient Boosting} & None (Original Data) & 99.0/99.50/99.69 & 99.00/99.50/99.69 & 99.0/99.50/99.69 & 99.0/99.50/99.69 \\
& Padding & 43.00/28.00/6.00 & 44.00/40.00/44.00 & 39.00/28.00/6.00 & 35.00/28.00/2.00 \\
& Padding + XORing & 42.00/28.00/6.00 & 44.00/39.00/43.00 & 39.00/28.00/6.00 & 34.00/27.00/2.00 \\
& Padding + Shifting & 42.00/27.00/6.00 & 43.00/40.00/29.00 & 39.00/27.00/6.00 & 34.00/26.00/2.00 \\
& Fragmentation & 52.00/30.00/33.00 & 47.00/45.00/72.00 & 42.00/30.00/33.00 & 39.00/29.00/36.00 \\
& Constant Size Padding & 48.00/26.00/16.00 & 44.00/19.00/59.00 & 42.00/26.00/16.00 & 38.00/21.00/17.00 \\
& Delay Randomization & 64.00/70.00/93.00 & 82.00/83.00/91.00 & 73.00/70.00/93.00 & 68.00/70.00/91.00 \\
\hline
\end{tabular}
}
\captionsetup{justification=centering}
\caption*{Performance Metrics: Acc: Accuracy, Prec: Precision, Rec: Recall, F1: F1 Score. Datasets: d1 (IoT-AD), d2 (IoT Sentinel), d3 (UNSW). Incremental Training and Fine-Tuning averages are shown for the Neural Network classifier.}
\end{table*}

\section{Numerical Results and Discussions} 
\subsection{Privacy-Preserving Experiments}
To evaluate our obfuscation framework, we used packet capture (pcap) files from three distinct public datasets. Each dataset underwent six different obfuscation techniques. We measured the performance of five classifiers, both before and after applying these techniques. The assessment focused on key performance metrics, including accuracy, precision, recall, and F1 score, to comprehensively understand the framework's impact. Additionally, incremental training and fine-tuning were conducted specifically on the neural network model with the obfuscated traffic to thoroughly evaluate the framework's effectiveness. The results, including those from incremental training and fine-tuning, are detailed in Table II, illustrating the framework's capability to enhance privacy and disrupt traffic analysis.

The results presented in Table II indicate that classifiers achieved high accuracy, precision, recall, and F1 scores, often exceeding 99\%, without obfuscation across all datasets. However, applying our obfuscation techniques led to a significant decline in these metrics. It is also observed from the results that despite adversaries retraining their models with obfuscated traffic through incremental training and fine-tuning, the performance metrics remained considerably lower than those for the original traffic, highlighting the effectiveness of our methods.

The introduction of 95\% confidence intervals and standard deviations into our analysis further reinforces the reliability and robustness of these results. For the original traffic, the 95\% confidence intervals (even at the lower bound) show that the classifiers perform consistently well, confirming that the model achieves strong classification results without obfuscation. However, after applying our obfuscation techniques, there is a notable decline in performance, even when looking at the upper bound of the 95\% confidence interval. This demonstrates that even in the best possible scenario for the obfuscated traffic, the classifiers still perform significantly worse than the original traffic, confirming the effectiveness of our framework. The lower standard deviations across both original and obfuscated traffic indicate consistent performance during stratified 10-fold cross-validation, confirming the reliability and stability of our framework in degrading classifiers' performance across all key metrics. Detailed results, including confidence intervals for the performance metrics, are provided in Table IV in the appendix for reference.

In scenarios where adversaries continually adapt their models over time, they may fine-tune or incrementally train their classifiers on obfuscated traffic to enhance detection accuracy. However, by randomly adjusting parameters such as padding size, fragment size, or delay timing, our framework effectively disrupts these efforts, making it extremely challenging for adversaries to identify consistent patterns. This dynamic approach ensures that even as adversaries attempt to exploit obfuscated data over time, the evolving nature of our techniques renders their models ineffective. This resilience is achieved through a combination of multiple obfuscation methods and the continuous random adjustment of their parameters.

Each dataset validated the efficacy of our techniques, showing a consistent pattern of reduced classifier performance after our obfuscation framework had been applied. This comprehensive evaluation underscores the capability of our framework to disrupt traffic analysis and enhance IoT privacy. 

\begin{table*}[ht]
\centering
\caption{Performance Metrics for Different Obfuscation Techniques}
\label{table:performance-metrics}
\begin{tabular}{
    |>{\raggedright}p{4cm}|
    >{\centering\arraybackslash}p{3cm}|
    >{\centering\arraybackslash}p{3cm}|
    >{\centering\arraybackslash}p{3cm}|
    >{\centering\arraybackslash}p{3cm}|
}
\hline
\rowcolor{lightgray}
\textbf{Obfuscation Technique} & \textbf{Execution Time (s)} & \textbf{Memory Used (MB)} & \textbf{Approx. CPU Usage (\%)} & \textbf{Bytes Added} \\
\hline
Padding & 0.002 & 0.017 & 0.001 & 127.078 \\
Padding and Shifting & 0.002 & 0.016 & 0.001 & 127.802 \\
Padding and XORing & 0.002 & 0.015 & 0.001 & 127.603 \\
Fragmentation & 0.003 & 0.019 & 0.002 & 34.371 \\
Constant Size Padding & 0.005 & 0.025 & 0.002 & 1,274.100 \\
Delay Randomization & 0.066 & 0.009 & 0.0002 & 0.000 \\
\hline
\end{tabular}
\captionsetup{justification=centering}
\caption*{Performance metrics for each obfuscation technique based on single packet processing.}
\end{table*}

\subsection{Trade-offs in Obfuscation Techniques}
Our experiments across the three datasets consistently demonstrate that while comprehensive obfuscation techniques effectively reduce the evaluation metrics of traffic analysis models, they could introduce trade-offs in terms of increased communication overhead, latency, and processing time.

\begin{itemize}
    \item \textbf{Padding:} For the padding technique, we randomly chose a byte size between 1 and 256 bytes to add to the packet payloads. This range was selected to obscure the original packet size effectively. During incremental training and fine-tuning, a smaller range of 1 to 128 bytes was used. This involves training the neural network model with the padded traffic, and then applying another layer of padding with the new range on the original traffic to test the model's effectiveness. Increasing the padding size beyond 256 bytes would further reduce the inference accuracy of the classifiers but at the cost of higher communication overhead due to increased data size.
    \item \textbf{Padding and XORing:} In the Padding and XORing technique, we applied a similar approach to padding, combined with XORing the payload with random bytes. The same ranges (1-256 bytes for initial obfuscation and 1-128 bytes for incremental and fine-tuning) were used. This dual-layer obfuscation method involved training the neural network with the initial obfuscated traffic and then applying new obfuscation parameters to the original traffic for testing. Further increasing the padding sizes would significantly impact communication overhead due to larger data sizes but would reduce the model's accuracy.
    \item \textbf{Padding and Shifting:} Padding and Shifting added random bytes to the payload and then shifted the entire payload. Similar to the other padding techniques, we used a range of 1-256 bytes for initial obfuscation and 1-128 bytes for incremental and fine-tuning. This process involved training the neural network with initially obfuscated traffic and then applying new shifting parameters to test the model. Increasing the range further would enhance obfuscation but also increase communication overhead.
    \item \textbf{Constant Size Padding:} Constant Size Padding involves padding packets to a uniform size, determined by the largest observed packet length. This constant size was dynamically adjusted as larger packet lengths were observed. In the incremental and fine-tuning phase, a random delay of 0.01 to 0.1 seconds was added. The neural network was trained with the constant size padding and then tested with new delay parameters. While this technique prevented size-based inference attacks, increasing the constant size and delay would reduce classifier accuracy but also increase both communication overhead and latency.
    \item \textbf{Fragmentation:} For the fragmentation technique, packets were randomly divided into two parts, complicating traffic analysis. In the incremental training and fine-tuning phase, an additional random delay of 0.01 to 0.1 seconds was introduced between fragments. The neural network was trained with the fragmented traffic without delay and then tested with new fragmentation and delay parameters. While this further reduced classifier accuracy, it also increased latency. Increasing the number of fragments and delay would further reduce inference accuracy but would significantly impact communication performance by increasing latency.
    \item \textbf{Delay Randomization:} Delay Randomization introduced random delays between 0.01 and 0.1 seconds for initial obfuscation. During incremental training and fine-tuning, the delay range was increased to between 0.01 and 0.2 seconds. This method involved training the neural network with the initial delay and then testing it with the new delay range. This effectively added temporal unpredictability, reducing inference accuracy. However, further increasing the delay range would enhance security but also significantly impact latency.
\end{itemize}

\subsection{Communication and System Performance Experiments}
We evaluated the impact of our obfuscation framework on communication and system performance, with numerical results in Table III. The key findings from these evaluations, underscoring the generally minimal overhead introduced by our framework, are summarized below:

\subsubsection{Execution Time} This refers to the time taken to process a single packet through each obfuscation technique. Our primary goal is to have a small execution time to ensure that real-time IoT systems can maintain their performance even when traffic obfuscation is applied. Techniques such as Padding, Padding and XORing, and Padding and Shifting are lightweight in terms of execution time (0.002 seconds) because they perform simpler transformations on packet structure. Delay Randomization, while effective at obfuscating traffic patterns by introducing temporal unpredictability, incurs higher execution time (0.066 seconds), which may impact real-time performance. 

\subsubsection{Memory Usage} This metric indicates the amount of memory consumed during the application of each obfuscation technique. Memory consumption varied significantly across the obfuscation techniques. Lower memory usage is important for resource-constrained IoT devices, which often have limited memory capacity. Padding and XORing consumes the least memory (0.015 MB), making it well-suited for devices with stringent memory limitations. On the other hand, constant-size padding (0.025 MB) requires higher memory consumption due to the need to pad packets to the largest observed size. While this technique is effective at preventing size-based inference attacks, it may not be ideal for memory-constrained devices. 

\subsubsection{CPU Usage} CPU utilization reflects the percentage of processor resources required to apply each obfuscation technique. Delay Randomization had the lowest CPU usage at 0.0002\%, indicating minimal impact on the processor. Conversely, Constant Size Padding and Fragmentation were the most CPU-intensive, utilizing 0.002\%, significantly increasing the processor load. These findings illustrate that while some techniques exert minimal pressure on the CPU, others can substantially increase the computational burden.
    
\subsubsection{Extra Bytes} This metric measures the additional bytes introduced by each obfuscation technique. In the context of IoT networks, minimizing communication overhead is crucial for preserving bandwidth, especially in low-power wide-area networks or other environments with limited data transmission capacity. The additional bytes introduced by each obfuscation technique to the original traffic were quantified to assess communication overhead. Fragmentation added the fewest bytes at 34.371, indicating minimal communication overhead. In stark contrast, Constant Size Padding resulted in the highest communication overhead, adding 1,274.100 bytes. This underscores the substantial variation in communication overhead associated with different techniques, reflecting the trade-off between the level of obfuscation and the added network load.

\section{Conclusion} 

In conclusion, our research introduces a comprehensive obfuscation framework that safeguards the privacy of IoT devices and user activities. By using six diverse obfuscation techniques, our framework effectively reduces the efficacy of traffic analysis attacks, demonstrated by a significant decline in the performance metrics of machine learning classifiers across multiple datasets. The thorough evaluation using three public datasets proves the robustness of our approach in various environments and its suitability for different traffic patterns and threats.

Our findings suggest that increasing the parameters of these techniques enhances their ability to disrupt traffic analysis, albeit with some communication and system overhead. This necessitates a strategic balance between privacy protection and system efficiency. The adoption of such comprehensive strategies is crucial in the evolving IoT landscape, where the rise of connected devices demands advanced measures for safeguarding data and privacy.

This study makes a significant contribution to IoT security, providing a scalable and effective solution to counter sophisticated traffic analysis methods. As IoT technology integrates more deeply into daily life, frameworks like ours will be vital for ensuring the confidentiality and integrity of personal and sensitive data, paving the way for future advancements in secure IoT environments.




\bibliography{obfuscation}

\begin{thebibliography}{10}
\providecommand{\url}[1]{#1}
\csname url@samestyle\endcsname
\providecommand{\newblock}{\relax}
\providecommand{\bibinfo}[2]{#2}
\providecommand{\BIBentrySTDinterwordspacing}{\spaceskip=0pt\relax}
\providecommand{\BIBentryALTinterwordstretchfactor}{4}
\providecommand{\BIBentryALTinterwordspacing}{\spaceskip=\fontdimen2\font plus
\BIBentryALTinterwordstretchfactor\fontdimen3\font minus \fontdimen4\font\relax}
\providecommand{\BIBforeignlanguage}[2]{{%
\expandafter\ifx\csname l@#1\endcsname\relax
\typeout{** WARNING: IEEEtran.bst: No hyphenation pattern has been}%
\typeout{** loaded for the language `#1'. Using the pattern for}%
\typeout{** the default language instead.}%
\else
\language=\csname l@#1\endcsname
\fi
#2}}
\providecommand{\BIBdecl}{\relax}
\BIBdecl

\bibitem{perera2022iot}
Y.~Perera, N.~Ahmed, S.~Kanhere, W.~Hu, and S.~Jha, ``Iot traffic obfuscation: Will it guarantee the privacy of your smart home?'' in \emph{ICC 2022-IEEE International Conference on Communications}.\hskip 1em plus 0.5em minus 0.4em\relax IEEE, 2022, pp. 2954--2959.

\bibitem{antonakakis2017understanding}
M.~Antonakakis, T.~April, M.~Bailey, M.~Bernhard, E.~Bursztein, J.~Cochran, Z.~Durumeric, J.~A. Halderman, L.~Invernizzi, M.~Kallitsis \emph{et~al.}, ``Understanding the mirai botnet,'' in \emph{26th USENIX security symposium (USENIX Security 17)}, 2017, pp. 1093--1110.

\bibitem{sivanathan2017characterizing}
A.~Sivanathan, D.~Sherratt, H.~H. Gharakheili, A.~Radford, C.~Wijenayake, A.~Vishwanath, and V.~Sivaraman, ``Characterizing and classifying iot traffic in smart cities and campuses,'' in \emph{2017 IEEE Conference on Computer Communications Workshops (INFOCOM WKSHPS)}.\hskip 1em plus 0.5em minus 0.4em\relax IEEE, 2017, pp. 559--564.

\bibitem{meidan2017profiliot}
Y.~Meidan, M.~Bohadana, A.~Shabtai, J.~D. Guarnizo, M.~Ochoa, N.~O. Tippenhauer, and Y.~Elovici, ``Profiliot: A machine learning approach for iot device identification based on network traffic analysis,'' in \emph{Proceedings of the symposium on applied computing}, 2017, pp. 506--509.

\bibitem{zhang2018homonit}
W.~Zhang, Y.~Meng, Y.~Liu, X.~Zhang, Y.~Zhang, and H.~Zhu, ``Homonit: Monitoring smart home apps from encrypted traffic,'' in \emph{Proceedings of the 2018 ACM SIGSAC Conference on Computer and Communications Security}, 2018, pp. 1074--1088.

\bibitem{wang2020fingerprinting}
C.~Wang, S.~Kennedy, H.~Li, K.~Hudson, G.~Atluri, X.~Wei, W.~Sun, and B.~Wang, ``Fingerprinting encrypted voice traffic on smart speakers with deep learning,'' in \emph{Proceedings of the 13th ACM Conference on Security and Privacy in Wireless and Mobile Networks}, 2020, pp. 254--265.

\bibitem{alshehri2020attacking}
A.~Alshehri, J.~Granley, and C.~Yue, ``Attacking and protecting tunneled traffic of smart home devices,'' in \emph{Proceedings of the Tenth ACM Conference on Data and Application Security and Privacy}, 2020, pp. 259--270.

\bibitem{acar2020peek}
A.~Acar, H.~Fereidooni, T.~Abera, A.~K. Sikder, M.~Miettinen, H.~Aksu, M.~Conti, A.-R. Sadeghi, and S.~Uluagac, ``Peek-a-boo: I see your smart home activities, even encrypted!'' in \emph{Proceedings of the 13th ACM Conference on Security and Privacy in Wireless and Mobile Networks}, 2020, pp. 207--218.

\bibitem{apthorpe2018keeping}
N.~Apthorpe, D.~Y. Huang, D.~Reisman, A.~Narayanan, and N.~Feamster, ``Keeping the smart home private with smart (er) iot traffic shaping,'' \emph{arXiv preprint arXiv:1812.00955}, 2018.

\bibitem{9723011}
D.~Chen, H.~Wang, N.~Zhang, X.~Nie, H.-N. Dai, K.~Zhang, and K.-K. Raymond~Choo, ``Privacy-preserving encrypted traffic inspection with symmetric cryptographic techniques in iot,'' \emph{IEEE Internet of Things Journal}, vol.~9, no.~18, pp. 17\,265--17\,279, 2022.

\bibitem{7845477}
S.~Chakrabarty, M.~John, and D.~W. Engels, ``Black routing and node obscuring in iot,'' in \emph{2016 IEEE 3rd World Forum on Internet of Things (WF-IoT)}, 2016, pp. 323--328.

\bibitem{REN2021105}
\BIBentryALTinterwordspacing
W.~Ren, X.~Tong, J.~Du, N.~Wang, S.~C. Li, G.~Min, Z.~Zhao, and A.~K. Bashir, ``Privacy-preserving using homomorphic encryption in mobile iot systems,'' \emph{Computer Communications}, vol. 165, pp. 105--111, 2021. [Online]. Available: \url{https://www.sciencedirect.com/science/article/pii/S0140366420319708}
\BIBentrySTDinterwordspacing

\bibitem{10.1145/2785956.2787502}
\BIBentryALTinterwordspacing
J.~Sherry, C.~Lan, R.~A. Popa, and S.~Ratnasamy, ``Blindbox: Deep packet inspection over encrypted traffic.''\hskip 1em plus 0.5em minus 0.4em\relax New York, NY, USA: Association for Computing Machinery, 2015. [Online]. Available: \url{https://doi.org/10.1145/2785956.2787502}
\BIBentrySTDinterwordspacing

\bibitem{194934}
\BIBentryALTinterwordspacing
C.~Lan, J.~Sherry, R.~A. Popa, S.~Ratnasamy, and Z.~Liu, ``Embark: Securely outsourcing middleboxes to the cloud,'' in \emph{13th USENIX Symposium on Networked Systems Design and Implementation (NSDI 16)}.\hskip 1em plus 0.5em minus 0.4em\relax Santa Clara, CA: USENIX Association, Mar. 2016, pp. 255--273. [Online]. Available: \url{https://www.usenix.org/conference/nsdi16/technical-sessions/presentation/lan}
\BIBentrySTDinterwordspacing

\bibitem{msadek2019iot}
N.~Msadek, R.~Soua, and T.~Engel, ``Iot device fingerprinting: Machine learning based encrypted traffic analysis,'' in \emph{2019 IEEE wireless communications and networking conference (WCNC)}.\hskip 1em plus 0.5em minus 0.4em\relax IEEE, 2019, pp. 1--8.

\bibitem{apthorpe2017spying}
N.~Apthorpe, D.~Reisman, S.~Sundaresan, A.~Narayanan, and N.~Feamster, ``Spying on the smart home: Privacy attacks and defenses on encrypted iot traffic,'' \emph{arXiv preprint arXiv:1708.05044}, 2017.

\bibitem{apthorpe2017smart}
N.~Apthorpe, D.~Reisman, and N.~Feamster, ``A smart home is no castle: Privacy vulnerabilities of encrypted iot traffic,'' \emph{arXiv preprint arXiv:1705.06805}, 2017.

\bibitem{10.1145/3559613.3563191}
\BIBentryALTinterwordspacing
A.~Engelberg and A.~Wool, ``Classification of encrypted iot traffic despite padding and shaping,'' in \emph{Proceedings of the 21st Workshop on Privacy in the Electronic Society}, ser. WPES'22.\hskip 1em plus 0.5em minus 0.4em\relax New York, NY, USA: Association for Computing Machinery, 2022, p. 1–13. [Online]. Available: \url{https://doi.org/10.1145/3559613.3563191}
\BIBentrySTDinterwordspacing

\bibitem{8538744}
A.~J. Pinheiro, J.~M. Bezerra, and D.~R. Campelo, ``Packet padding for improving privacy in consumer iot,'' in \emph{2018 IEEE Symposium on Computers and Communications (ISCC)}, 2018, pp. 00\,925--00\,929.

\bibitem{datta2018developer}
T.~Datta, N.~Apthorpe, and N.~Feamster, ``A developer-friendly library for smart home iot privacy-preserving traffic obfuscation,'' in \emph{Proceedings of the 2018 workshop on IoT security and privacy}, 2018, pp. 43--48.

\bibitem{xiong2018defending}
S.~Xiong, A.~D. Sarwate, and N.~B. Mandayam, ``Defending against packet-size side-channel attacks in iot networks,'' in \emph{2018 IEEE International Conference on Acoustics, Speech and Signal Processing (ICASSP)}.\hskip 1em plus 0.5em minus 0.4em\relax IEEE, 2018, pp. 2027--2031.

\bibitem{9825227}
F.~Shen, S.~Zhang, Y.~Liu, and Z.~Yang, ``A survey of traffic obfuscation technology for smart home,'' in \emph{2022 International Wireless Communications and Mobile Computing (IWCMC)}, 2022, pp. 997--1002.

\bibitem{9750450}
B.~Chen, Y.~Liu, S.~Zhang, J.~Chen, and Z.~Han, ``A survey on smart home privacy data protection technology,'' in \emph{2021 IEEE Sixth International Conference on Data Science in Cyberspace (DSC)}, 2021, pp. 583--590.

\bibitem{6234422}
K.~P. Dyer, S.~E. Coull, T.~Ristenpart, and T.~Shrimpton, ``Peek-a-boo, i still see you: Why efficient traffic analysis countermeasures fail,'' in \emph{2012 IEEE Symposium on Security and Privacy}, 2012, pp. 332--346.

\bibitem{9203848}
A.~J. Pinheiro, P.~Freitas~de Araujo-Filho, J.~de~M.~Bezerra, and D.~R. Campelo, ``Adaptive packet padding approach for smart home networks: A tradeoff between privacy and performance,'' \emph{IEEE Internet of Things Journal}, vol.~8, no.~5, pp. 3930--3938, 2021.

\bibitem{5961736}
F.~Zhang, W.~He, and X.~Liu, ``Defending against traffic analysis in wireless networks through traffic reshaping,'' in \emph{2011 31st International Conference on Distributed Computing Systems}, 2011, pp. 593--602.

\bibitem{10.1145/3374664.3375723}
\BIBentryALTinterwordspacing
A.~Alshehri, J.~Granley, and C.~Yue, ``Attacking and protecting tunneled traffic of smart home devices,'' in \emph{Proceedings of the Tenth ACM Conference on Data and Application Security and Privacy}, ser. CODASPY '20.\hskip 1em plus 0.5em minus 0.4em\relax New York, NY, USA: Association for Computing Machinery, 2020, p. 259–270. [Online]. Available: \url{https://doi.org/10.1145/3374664.3375723}
\BIBentrySTDinterwordspacing

\bibitem{hussain2021dark}
A.~M. Hussain, G.~Oligeri, and T.~Voigt, ``The dark (and bright) side of iot: Attacks and countermeasures for identifying smart home devices and services,'' in \emph{Security, Privacy, and Anonymity in Computation, Communication, and Storage: SpaCCS 2020 International Workshops, Nanjing, China, December 18-20, 2020, Proceedings 13}.\hskip 1em plus 0.5em minus 0.4em\relax Springer, 2021, pp. 122--136.

\bibitem{9322070}
N.~Prates, A.~Vergütz, R.~T. Macedo, A.~Santos, and M.~Nogueira, ``A defense mechanism for timing-based side-channel attacks on iot traffic,'' in \emph{GLOBECOM 2020 - 2020 IEEE Global Communications Conference}, 2020, pp. 1--6.

\bibitem{8730787}
I.~Hafeez, M.~Antikainen, and S.~Tarkoma, ``Protecting iot-environments against traffic analysis attacks with traffic morphing,'' in \emph{2019 IEEE International Conference on Pervasive Computing and Communications Workshops (PerCom Workshops)}, 2019, pp. 196--201.

\bibitem{zhu2021smart}
Q.~Zhu, C.~Yang, Y.~Zheng, J.~Ma, H.~Li, J.~Zhang, and J.~Shao, ``Smart home: Keeping privacy based on air-padding,'' \emph{IET Information Security}, vol.~15, no.~2, pp. 156--168, 2021.

\bibitem{9609087}
J.~Brahma and D.~Sadhya, ``Preserving contextual privacy for smart home iot devices with dynamic traffic shaping,'' \emph{IEEE Internet of Things Journal}, vol.~9, no.~13, pp. 11\,434--11\,441, 2022.

\bibitem{apthorpe2017closing}
N.~Apthorpe, D.~Reisman, and N.~Feamster, ``Closing the blinds: Four strategies for protecting smart home privacy from network observers,'' \emph{arXiv preprint arXiv:1705.06809}, 2017.

\bibitem{8704324}
R.~Xu, Q.~Zeng, L.~Zhu, H.~Chi, X.~Du, and M.~Guizani, ``Privacy leakage in smart homes and its mitigation: Ifttt as a case study,'' \emph{IEEE Access}, vol.~7, pp. 63\,457--63\,471, 2019.

\bibitem{9464026}
C.~Duan, S.~Zhang, J.~Yang, Z.~Wang, Y.~Yang, and J.~Li, ``Pinball: Universal and robust signature extraction for smart home devices,'' in \emph{2021 IFIP/IEEE International Symposium on Integrated Network Management (IM)}, 2021, pp. 1--9.

\bibitem{zhang2023novel}
S.~Zhang, F.~Shen, Y.~Liu, Z.~Yang, and X.~Lv, ``A novel traffic obfuscation technology for smart home,'' \emph{Electronics}, vol.~12, no.~16, p. 3477, 2023.

\bibitem{trimananda2019pingpong}
R.~Trimananda, J.~Varmarken, A.~Markopoulou, and B.~Demsky, ``Pingpong: Packet-level signatures for smart home device events,'' \emph{arXiv preprint arXiv:1907.11797}, 2019.

\bibitem{danso2023transferability}
P.~K. Danso, S.~Dadkhah, E.~C.~P. Neto, A.~Zohourian, H.~Molyneaux, R.~Lu, and A.~A. Ghorbani, ``Transferability of machine learning algorithm for iot device profiling and identification,'' \emph{IEEE Internet of Things Journal}, 2023.

\bibitem{zahan2023iot}
H.~Zahan, M.~W. Al~Azad, I.~Ali, and S.~Mastorakis, ``Iot-ad: A framework to detect anomalies among interconnected iot devices,'' \emph{IEEE Internet of Things Journal}, vol.~11, no.~1, pp. 478--489, 2023.

\bibitem{miettinen2017iot}
M.~Miettinen, S.~Marchal, I.~Hafeez, N.~Asokan, A.-R. Sadeghi, and S.~Tarkoma, ``Iot sentinel: Automated device-type identification for security enforcement in iot,'' in \emph{2017 IEEE 37th international conference on distributed computing systems (ICDCS)}.\hskip 1em plus 0.5em minus 0.4em\relax IEEE, 2017, pp. 2177--2184.

\bibitem{sivanathan2018classifying}
A.~Sivanathan, H.~H. Gharakheili, F.~Loi, A.~Radford, C.~Wijenayake, A.~Vishwanath, and V.~Sivaraman, ``Classifying iot devices in smart environments using network traffic characteristics,'' \emph{IEEE Transactions on Mobile Computing}, vol.~18, no.~8, pp. 1745--1759, 2018.

\end{thebibliography}

\bibliographystyle{IEEEtran}

\appendix


\begin{table*}[ht]
\centering
\caption{Performance Metrics for Various Obfuscation Techniques Across Three Datasets (IoT-AD, IoT Sentinel, UNSW)}
\label{table:performance-metrics_3}
\resizebox{1\textwidth}{!}{%
\begin{tabular}{
    |>{\raggedright}p{2cm}|
    >{\raggedright}p{2.7cm}|
    >{\centering\arraybackslash}p{4.8cm}|
    >{\centering\arraybackslash}p{4.8cm}|
    >{\centering\arraybackslash}p{4.8cm}|
    >{\centering\arraybackslash}p{4.8cm}|
}
\hline
\rowcolor{lightgray}
\textbf{Classifier} & \textbf{Obfuscation Technique} & \textbf{Accuracy (Mean ± SD, 95\% CI for d1 / d2 / d3)} & \textbf{Precision (Mean ± SD, 95\% CI for d1 / d2 / d3)} & \textbf{Recall (Mean ± SD, 95\% CI for d1 / d2 / d3)} & \textbf{F1 Score (Mean ± SD, 95\% CI for d1 / d2 / d3)} \\
\hline
\multirow{10}{*}{Neural Network} 
& None (Original Data) & 95.15 ± 0.01043 (94.4 - 95.89) / 92.19 ± 0.00727 (91.67 - 92.71) / 98.54 ± 0.002 (98.4 - 98.69) 
& 94.35 ± 0.01185 (93.51 - 95.2) / 92.71 ± 0.00748 (92.17 - 93.24) / 98.62 ± 0.00195 (98.48 - 98.76) 
& 92.33 ± 0.02604 (90.46 - 94.19) / 92.37 ± 0.00705 (91.87 - 92.88) / 98.53 ± 0.00201 (98.39 - 98.68) 
& 93.15 ± 0.0197 (91.74 - 94.56) / 92.48 ± 0.00721 (91.97 - 93.0) / 98.57 ± 0.00197 (98.43 - 98.71) \\

& Padding & 29.74 ± 0.0 (29.74 - 29.74) / 24.81 ± 0.02798 (22.81 - 26.82) / 6.7 ± 0.05315 (2.9 - 10.51) 
& 7.97 ± 0.06066 (3.63 - 12.31) / 13.05 ± 0.12833 (3.87 - 22.23) / 16.47 ± 0.05183 (12.76 - 20.18) 
& 20.0 ± 9E-05 (20.0 - 20.01) / 22.21 ± 0.02971 (20.09 - 24.34) / 16.61 ± 0.0017 (16.48 - 16.73) 
& 9.2 ± 0.00102 (9.13 - 9.28) / 10.88 ± 0.04177 (7.89 - 13.87) / 2.4 ± 0.02131 (0.88 - 3.93) \\

& Padding + XORing & 26.59 ± 0.07983 (20.88 - 32.3) / 24.2 ± 0.00977 (23.5 - 24.9) / 7.13 ± 0.04587 (3.85 - 10.41) 
& 7.41 ± 0.06292 (2.9 - 11.91) / 26.16 ± 0.14687 (15.65 - 36.66) / 12.6 ± 0.06302 (8.09 - 17.1) 
& 18.2 ± 0.05426 (14.31 - 22.08) / 21.59 ± 0.01029 (20.86 - 22.33) / 18.8 ± 0.04736 (15.41 - 22.19) 
& 8.31 ± 0.02245 (6.71 - 9.92) / 10.49 ± 0.01826 (9.18 - 11.8) / 3.89 ± 0.04502 (0.67 - 7.11) \\

& Padding + Shifting & 29.25 ± 0.0 (29.25 - 29.25) / 23.13 ± 0.01585 (22.0 - 24.27) / 5.01 ± 0.00198 (4.87 - 5.16) 
& 5.85 ± 1E-05 (5.85 - 5.85) / 21.3 ± 0.17619 (8.69 - 33.9) / 1.32 ± 0.01333 (0.36 - 2.27) 
& 20.0 ± 0.0 (20.0 - 20.0) / 21.33 ± 0.01125 (20.52 - 22.13) / 16.24 ± 0.01292 (15.31 - 17.16) 
& 9.05 ± 1E-05 (9.05 - 9.05) / 10.06 ± 0.02128 (8.53 - 11.58) / 1.84 ± 0.00588 (1.42 - 2.26) \\

& Fragmentation & 35.79 ± 0.0 (35.79 - 35.79) / 19.46 ± 0.02738 (17.5 - 21.42) / 5.23 ± 0.00137 (5.13 - 5.33) 
& 26.91 ± 0.0 (26.91 - 26.91) / 5.94 ± 0.06053 (1.61 - 10.27) / 8.71 ± 0.10912 (0.91 - 16.52) 
& 21.09 ± 0.0 (21.09 - 21.09) / 20.96 ± 0.02888 (18.9 - 23.03) / 16.72 ± 0.00094 (16.66 - 16.79) 
& 12.34 ± 0.0 (12.34 - 12.34) / 7.88 ± 0.03981 (5.03 - 10.73) / 1.75 ± 0.00186 (1.61 - 1.88) \\

& Constant Size Padding & 32.46 ± 0.00077 (32.4 - 32.52) / 25.39 ± 0.02803 (23.38 - 27.39) / 11.16 ± 0.01561 (10.05 - 12.28) 
& 24.68 ± 0.02566 (22.85 - 26.52) / 9.35 ± 0.05649 (5.3 - 13.39) / 18.06 ± 0.05773 (13.93 - 22.18) 
& 21.75 ± 0.00042 (21.72 - 21.78) / 22.86 ± 0.03078 (20.66 - 25.06) / 20.75 ± 0.02291 (19.11 - 22.39) 
& 12.49 ± 0.00074 (12.44 - 12.55) / 11.76 ± 0.03657 (9.14 - 14.38) / 10.26 ± 0.01914 (8.89 - 11.63) \\

& Delay Randomization & 58.25 ± 0.08805 (51.95 - 64.55) / 71.1 ± 0.01561 (69.98 - 72.21) / 46.79 ± 0.0451 (43.56 - 50.01) 
& 55.18 ± 0.06598 (50.46 - 59.9) / 74.85 ± 0.02236 (73.25 - 76.44) / 58.28 ± 0.09698 (51.34 - 65.22) 
& 52.5 ± 0.06086 (48.15 - 56.86) / 72.96 ± 0.01433 (71.94 - 73.99) / 49.22 ± 0.04917 (45.7 - 52.74) 
& 50.24 ± 0.07898 (44.59 - 55.89) / 71.0 ± 0.01873 (69.66 - 72.34) / 41.4 ± 0.06235 (36.94 - 45.86) \\


\hline

\multirow{7}{*}{Random Forest} 
& None (Original Data) 
& 99.59 ± 0.00231 (99.42 - 99.75) / 99.88 ± 0.00098 (99.81 - 99.95) / 99.97 ± 0.00009 (99.96 - 99.98) 
& 99.41 ± 0.00386 (99.14 - 99.69) / 99.88 ± 0.00102 (99.81 - 99.95) / 99.97 ± 0.00008 (99.97 - 99.98) 
& 99.41 ± 0.00434 (99.10 - 99.72) / 99.88 ± 0.001 (99.81 - 99.95) / 99.96 ± 0.00018 (99.94 - 99.97) 
& 99.41 ± 0.00367 (99.14 - 99.67) / 99.88 ± 0.00101 (99.81 - 99.95) / 99.96 ± 0.00012 (99.96 - 99.97) \\

& Padding 
& 65.84 ± 0.00586 (65.42 - 66.26) / 54.03 ± 0.04215 (51.01 - 57.04) / 46.12 ± 0.05075 (42.48 - 49.75) 
& 53.79 ± 0.13322 (44.26 - 63.32) / 58.39 ± 0.07069 (53.34 - 63.45) / 63.34 ± 0.03188 (61.06 - 65.62) 
& 51.25 ± 0.00398 (50.96 - 51.53) / 54.55 ± 0.04731 (51.17 - 57.94) / 51.43 ± 0.05267 (47.66 - 55.30) 
& 47.06 ± 0.00311 (46.84 - 47.28) / 53.57 ± 0.06010 (49.27 - 57.87) / 37.61 ± 0.06859 (32.70 - 42.51) \\

& Padding + XORing & 64.94 ± 0.00585 (64.52 - 65.35) / 53.32 ± 0.04033 (50.44 - 56.2) / 46.16 ± 0.0514 (42.48 - 49.83) 
& 53.37 ± 0.13081 (44.01 - 62.73) / 57.51 ± 0.06813 (52.64 - 62.39) / 63.78 ± 0.03162 (61.51 - 66.04) 
& 51.37 ± 0.00344 (51.13 - 51.62) / 53.86 ± 0.04548 (50.61 - 57.12) / 51.49 ± 0.05347 (47.66 - 55.31) 
& 46.82 ± 0.00279 (46.62 - 47.02) / 52.89 ± 0.05852 (48.71 - 57.08) / 37.75 ± 0.06931 (32.79 - 42.71) \\

& Padding + Shifting & 65.03 ± 0.00491 (64.68 - 65.39) / 52.41 ± 0.0419 (49.41 - 55.41) / 46.15 ± 0.05161 (42.46 - 49.84)
& 51.43 ± 0.0921 (44.84 - 58.02) / 57.04 ± 0.06943 (52.07 - 62.00) / 63.37 ± 0.03169 (61.06 - 65.67)
& 51.43 ± 0.0033 (51.2 - 51.67) / 52.61 ± 0.04658 (49.28 - 55.95) / 51.46 ± 0.0537 (47.62 - 55.3)
& 46.87 ± 0.00264 (46.68 - 47.06) / 51.99 ± 0.05951 (47.74 - 56.25) / 37.68 ± 0.06938 (32.7 - 42.64) \\

& Fragmentation & 69.34 ± 0.00302 (69.12 - 69.55) / 51.1 ± 0.027 (49.16 - 53.03) / 49.07 ± 0.03855 (46.31 - 51.82)
& 56.98 ± 0.13528 (47.3 - 66.66) / 51.87 ± 0.04533 (48.63 - 55.12) / 48.3 ± 0.01823 (47.0 - 49.61)
& 52.17 ± 0.00171 (52.05 - 52.29) / 50.03 ± 0.02617 (48.16 - 51.91) / 56.7 ± 0.03245 (54.38 - 59.02)
& 48.47 ± 0.00429 (48.17 - 48.78) / 45.1 ± 0.03284 (42.76 - 47.45) / 43.69 ± 0.03809 (40.97 - 46.42) \\

& Constant Size Padding & 63.89 ± 0.0119 (63.03 - 64.74) / 50.25 ± 0.07406 (44.95 - 55.55) / 41.08 ± 0.07805 (35.5 - 46.66)
& 40.86 ± 0.01526 (39.77 - 41.95) / 58.37 ± 0.11309 (50.28 - 66.46) / 49.77 ± 0.11956 (41.22 - 58.32)
& 50.77 ± 0.00611 (50.33 - 51.2) / 50.69 ± 0.07995 (44.97 - 56.41) / 42.7 ± 0.08405 (36.69 - 48.71)
& 44.58 ± 0.01108 (43.78 - 45.37) / 48.23 ± 0.10284 (40.87 - 55.59) / 33.77 ± 0.10674 (26.14 - 41.41) \\

& Delay Randomization & 81.21 ± 0.02262 (79.59 - 82.83) / 72.24 ± 0.01802 (70.95 - 73.53) / 96.4 ± 0.00474 (96.06 - 96.74)
& 86.27 ± 0.01157 (85.44 - 87.1) / 79.04 ± 0.01297 (78.11 - 79.97) / 95.87 ± 0.00487 (95.53 - 96.22)
& 80.85 ± 0.05107 (77.2 - 84.5) / 74.46 ± 0.01809 (73.17 - 75.76) / 95.62 ± 0.00564 (95.21 - 96.02)
& 80.44 ± 0.05177 (76.74 - 84.14) / 72.32 ± 0.01781 (71.04 - 73.59) / 95.71 ± 0.00531 (95.33 - 96.09) \\
\hline

\multirow{7}{*}{K-NN} 
& None (Original Data) 
& 92.50 ± 0.00904 (91.86 - 93.15) / 90.94 ± 0.00447 (90.62 - 91.26) / 97.20 ± 0.0011 (97.12 - 97.28) 
& 90.44 ± 0.00992 (89.73 - 91.15) / 91.21 ± 0.00529 (90.84 - 91.59) / 97.32 ± 0.001 (97.24 - 97.39) 
& 89.95 ± 0.01472 (88.9 - 91.01) / 90.92 ± 0.00482 (90.57 - 91.26) / 96.99 ± 0.00165 (96.87 - 97.11) 
& 90.13 ± 0.01176 (89.29 - 90.97) / 91.04 ± 0.00493 (90.68 - 91.39) / 97.15 ± 0.00122 (97.06 - 97.24) \\

& Padding 
& 29.74 ± 1e-6 (29.74 - 29.74) / 22.74 ± 1e-6 (22.74 - 22.74) / 4.95 ± 1e-6 (4.95 - 4.95) 
& 5.96 ± 7E-05 (5.95 - 5.96) / 4.55 ± 1e-6 (4.55 - 4.55) / 17.49 ± 1e-6 (17.49 - 17.49) 
& 20.0 ± 1e-6 (20.0 - 20.0) / 20.0 ± 1e-6 (20.0 - 20.0) / 16.67 ± 1e-6 (16.67 - 16.67) 
& 9.18 ± 8E-05 (9.17 - 9.18) / 7.41 ± 1e-6 (7.41 - 7.41) / 1.58 ± 1E-05 (1.58 - 1.58) \\

& Padding + XORing & 29.25 ± 1e-6 (29.25 - 29.25) / 24.95 ± 1e-6 (24.95 - 24.95) / 4.99 ± 1E-05 (4.99 - 5.0) 
& 5.86 ± 6E-05 (5.85 - 5.86) / 44.65 ± 1e-6 (44.65 - 44.65) / 7.84 ± 0.00067 (7.79 - 7.89) 
& 20.0 ± 1e-6 (20.0 - 20.0) / 22.38 ± 1e-6 (22.38 - 22.38) / 16.71 ± 1E-05 (16.71 - 16.71) 
& 9.06 ± 7E-05 (9.05 - 9.07) / 11.98 ± 1e-6 (11.98 - 11.98) / 1.66 ± 1E-05 (1.66 - 1.66) \\

& Padding + Shifting & 29.25 ± 1e-6 (29.25 - 29.25) / 23.83 ± 1e-6 (23.83 - 23.83) / 4.95 ± 1e-6 (4.95 - 4.95)
& 5.86 ± 7E-05 (5.85 - 5.86) / 16.47 ± 1e-6 (16.47 - 16.47) / 0.82 ± 1e-6 (0.82 - 0.82)
& 20.0 ± 1e-6 (20.0 - 20.0) / 21.51 ± 1e-6 (21.51 - 21.51) / 16.67 ± 1e-6 (16.67 - 16.67)
& 9.06 ± 8E-05 (9.05 - 9.07) / 10.15 ± 1e-6 (10.15 - 10.15) / 1.57 ± 1E-05 (1.57 - 1.57) \\

& Fragmentation & 35.55 ± 1e-6 (35.55 - 35.55) / 18.55 ± 1e-6 (18.55 - 18.55) / 5.15 ± 1e-6 (5.15 - 5.15)
& 26.89 ± 1e-6 (26.89 - 26.89) / 3.71 ± 1e-6 (3.71 - 3.71) / 0.86 ± 1e-6 (0.86 - 0.86)
& 20.95 ± 1e-6 (20.95 - 20.95) / 20.0 ± 1e-6 (20.0 - 20.0) / 16.67 ± 1e-6 (16.67 - 16.67)
& 12.07 ± 1e-6 (12.07 - 12.07) / 6.26 ± 1e-6 (6.26 - 6.26) / 1.63 ± 1e-6 (1.63 - 1.63) \\

& Constant Size Padding & 31.03 ± 0.0038 (30.76 - 31.31) / 22.74 ± 1e-6 (22.74 - 22.74) / 8.26 ± 0.00025 (8.24 - 8.28)
& 24.25 ± 0.01768 (22.98 - 25.51) / 4.89 ± 1e-6 (4.89 - 4.89) / 19.67 ± 0.00564 (19.26 - 20.07)
& 20.97 ± 0.00207 (20.82 - 21.12) / 20.0 ± 1e-6 (20.0 - 20.0) / 18.1 ± 0.00028 (18.08 - 18.12)
& 11.08 ± 0.00371 (10.82 - 11.35) / 7.86 ± 1e-6 (7.86 - 7.86) / 7.04 ± 0.00032 (7.02 - 7.06) \\

& Delay Randomization & 75.46 ± 0.00235 (75.29 - 75.62) / 66.53 ± 0.00283 (66.33 - 66.73) / 80.14 ± 0.00265 (79.95 - 80.33)
& 72.81 ± 0.00372 (72.55 - 73.08) / 68.96 ± 0.00233 (68.8 - 69.13) / 75.26 ± 0.00285 (75.06 - 75.46)
& 73.83 ± 0.00346 (73.58 - 74.08) / 67.93 ± 0.00253 (67.75 - 68.12) / 75.05 ± 0.00206 (74.91 - 75.2)
& 73.05 ± 0.00298 (72.84 - 73.27) / 66.77 ± 0.00361 (66.51 - 67.02) / 74.78 ± 0.00249 (74.6 - 74.96) \\
\hline

\multirow{7}{*}{Decision Tree} 
& None (Original Data) 
& 99.18 ± 0.00412 (98.89 - 99.48) / 99.77 ± 0.00103 (99.7 - 99.84) / 99.93 ± 0.00017 (99.92 - 99.94) 
& 98.78 ± 0.00593 (98.36 - 99.21) / 99.77 ± 0.00105 (99.69 - 99.84) / 99.92 ± 0.0004 (99.89 - 99.95) 
& 98.83 ± 0.00601 (98.4 - 99.26) / 99.77 ± 0.00109 (99.69 - 99.85) / 99.92 ± 0.00024 (99.91 - 99.94) 
& 98.80 ± 0.00547 (98.41 - 99.19) / 99.77 ± 0.00107 (99.69 - 99.84) / 99.92 ± 0.00028 (99.9 - 99.94) \\

& Padding 
& 51.90 ± 0.11738 (43.50 - 60.29) / 48.07 ± 0.04732 (44.69 - 51.46) / 39.74 ± 0.01821 (38.44 - 41.04) 
& 45.26 ± 0.07768 (39.70 - 50.81) / 48.50 ± 0.07227 (43.33 - 53.67) / 51.12 ± 0.05963 (46.86 - 55.39) 
& 44.40 ± 0.06134 (40.02 - 48.79) / 49.02 ± 0.04912 (45.50 - 52.53) / 41.68 ± 0.02233 (40.08 - 43.28) 
& 39.37 ± 0.06938 (34.41 - 44.33) / 46.68 ± 0.05142 (43.01 - 50.36) / 30.37 ± 0.02185 (28.81 - 31.93) \\

& Padding + XORing & 51.39 ± 0.11231 (43.36 - 59.42) / 47.39 ± 0.04753 (43.99 - 50.79) / 39.72 ± 0.01817 (38.42 - 41.02) 
& 43.52 ± 0.03776 (40.82 - 46.22) / 47.43 ± 0.07427 (42.12 - 52.74) / 49.09 ± 0.05791 (44.95 - 53.23) 
& 44.61 ± 0.05986 (40.33 - 48.89) / 48.4 ± 0.04935 (44.87 - 51.93) / 41.7 ± 0.02199 (40.13 - 43.27) 
& 39.32 ± 0.06657 (34.56 - 44.08) / 46.01 ± 0.0523 (42.27 - 49.75) / 30.41 ± 0.02127 (28.89 - 31.93) \\

& Padding + Shifting & 51.14 ± 0.11479 (42.92 - 59.35) / 46.58 ± 0.04607 (43.28 - 49.87) / 39.67 ± 0.01817 (38.37 - 40.97)
& 44.88 ± 0.07587 (39.45 - 50.3) / 47.06 ± 0.0739 (41.77 - 52.34) / 48.72 ± 0.06474 (44.08 - 53.35)
& 44.42 ± 0.06118 (40.05 - 48.8) / 47.28 ± 0.04903 (43.77 - 50.78) / 41.67 ± 0.02236 (40.08 - 43.28)
& 39.11 ± 0.06924 (34.15 - 44.06) / 45.07 ± 0.05164 (41.37 - 48.76) / 30.5 ± 0.02157 (28.95 - 32.04) \\

& Fragmentation & 58.03 ± 0.0893 (51.64 - 64.42) / 56.45 ± 0.03518 (53.93 - 58.97) / 45.25 ± 0.02832 (43.22 - 47.27)
& 44.14 ± 0.04045 (41.24 - 47.03) / 57.83 ± 0.02933 (55.73 - 59.93) / 33.04 ± 0.01748 (31.79 - 34.29)
& 46.7 ± 0.05 (43.13 - 50.28) / 54.78 ± 0.03873 (52.01 - 57.55) / 43.1 ± 0.01569 (41.97 - 44.22)
& 41.72 ± 0.05145 (38.04 - 45.4) / 50.27 ± 0.05521 (46.33 - 54.22) / 34.94 ± 0.0168 (33.74 - 36.14) \\

& Constant Size Padding & 54.02 ± 0.10922 (46.21 - 61.84) / 33.9 ± 0.06372 (29.34 - 38.46) / 31.87 ± 0.02282 (30.24 - 33.5)
& 38.6 ± 0.04093 (35.67 - 41.53) / 28.74 ± 0.11235 (20.71 - 36.78) / 20.75 ± 0.04771 (17.33 - 24.16)
& 45.86 ± 0.05826 (41.69 - 50.03) / 35.8 ± 0.06394 (31.23 - 40.38) / 30.96 ± 0.01454 (29.92 - 32.0)
& 38.78 ± 0.06779 (33.93 - 43.63) / 28.62 ± 0.08562 (22.49 - 34.74) / 20.2 ± 0.0263 (18.32 - 22.08) \\

& Delay Randomization & 68.49 ± 0.02027 (67.04 - 69.94) / 78.45 ± 0.0347 (75.97 - 80.94) / 83.14 ± 0.02113 (81.63 - 84.65)
& 67.4 ± 0.05612 (63.39 - 71.42) / 80.66 ± 0.03761 (77.97 - 83.35) / 79.52 ± 0.02566 (77.69 - 81.36)
& 60.58 ± 0.02015 (59.14 - 62.02) / 78.92 ± 0.03973 (76.08 - 81.77) / 79.34 ± 0.01632 (78.17 - 80.51)
& 59.22 ± 0.01597 (58.07 - 60.36) / 78.44 ± 0.03883 (75.66 - 81.22) / 77.86 ± 0.02364 (76.17 - 79.55) \\
\hline

\multirow{7}{*}{Gradient Boosting} 
& None (Original Data) 
& 99.52 ± 0.00298 (99.30 - 99.73) / 99.65 ± 0.00139 (99.55 - 99.75) / 99.67 ± 0.00039 (99.64 - 99.7) 
& 99.19 ± 0.00563 (98.78 - 99.59) / 99.66 ± 0.00141 (99.56 - 99.76) / 99.68 ± 0.00038 (99.65 - 99.71) 
& 99.44 ± 0.00432 (99.13 - 99.75) / 99.64 ± 0.00147 (99.54 - 99.75) / 99.51 ± 0.00079 (99.46 - 99.57) 
& 99.31 ± 0.00428 (99.00 - 99.61) / 99.65 ± 0.00144 (99.55 - 99.75) / 99.60 ± 0.00054 (99.56 - 99.63) \\

& Padding 
& 45.47 ± 0.04115 (42.52 - 48.41) / 33.03 ± 0.03304 (30.67 - 35.4) / 6.84 ± 0.01274 (5.93 - 7.75) 
& 43.28 ± 0.01351 (42.32 - 44.25) / 41.29 ± 0.02952 (39.17 - 43.4) / 40.51 ± 0.04719 (37.14 - 43.89) 
& 40.51 ± 0.02198 (38.94 - 42.08) / 31.75 ± 0.03402 (29.32 - 34.19) / 18.23 ± 0.01071 (17.46 - 19.0) 
& 35.97 ± 0.02574 (34.13 - 37.82) / 31.36 ± 0.0328 (29.02 - 33.71) / 4.50 ± 0.01421 (3.48 - 5.51) \\

& Padding + XORing & 44.44 ± 0.04223 (41.42 - 47.46) / 32.58 ± 0.04008 (29.71 - 35.45) / 6.89 ± 0.01287 (5.97 - 7.81) 
& 42.13 ± 0.01663 (40.94 - 43.32) / 40.3 ± 0.03404 (37.86 - 42.73) / 28.22 ± 0.05779 (24.08 - 32.35) 
& 40.37 ± 0.02294 (38.72 - 42.01) / 31.38 ± 0.0412 (28.43 - 34.33) / 18.27 ± 0.01082 (17.5 - 19.05) 
& 35.54 ± 0.02665 (33.63 - 37.45) / 30.82 ± 0.03733 (28.15 - 33.49) / 4.58 ± 0.01437 (3.55 - 5.6) \\

& Padding + Shifting & 44.38 ± 0.041 (41.44 - 47.31) / 31.32 ± 0.03803 (28.6 - 34.04) / 6.84 ± 0.01286 (5.92 - 7.76)
& 42.78 ± 0.01445 (41.75 - 43.81) / 39.98 ± 0.03374 (37.57 - 42.39) / 28.72 ± 0.06878 (23.8 - 33.64)
& 40.32 ± 0.0223 (38.72 - 41.92) / 30.12 ± 0.03933 (27.31 - 32.93) / 18.24 ± 0.01081 (17.46 - 19.01)
& 35.58 ± 0.02602 (33.72 - 37.44) / 29.34 ± 0.03616 (26.75 - 31.92) / 4.51 ± 0.01432 (3.48 - 5.53) \\

& Fragmentation & 54.23 ± 0.04349 (51.12 - 57.34) / 31.01 ± 0.02456 (29.25 - 32.77) / 29.55 ± 0.06016 (25.35 - 33.83)
& 48.32 ± 0.06007 (44.02 - 52.61) / 40.48 ± 0.02804 (38.48 - 42.49) / 46.28 ± 0.11207 (38.26 - 54.3)
& 43.49 ± 0.02492 (41.7 - 45.27) / 31.62 ± 0.02592 (29.77 - 33.48) / 39.67 ± 0.05724 (35.57 - 43.76)
& 40.63 ± 0.0253 (38.82 - 42.43) / 27.34 ± 0.02758 (25.37 - 29.31) / 30.73 ± 0.07544 (25.34 - 36.13) \\

& Constant Size Padding & 47.54 ± 0.00568 (47.14 - 47.95) / 17.51 ± 0.04813 (14.07 - 20.96) / 15.25 ± 0.00328 (15.02 - 15.48)
& 43.77 ± 0.00663 (43.29 - 44.24) / 17.34 ± 0.09387 (10.63 - 24.06) / 42.83 ± 0.06207 (38.39 - 47.27)
& 42.06 ± 0.00318 (41.83 - 42.28) / 17.35 ± 0.04982 (13.79 - 20.92) / 25.76 ± 0.0025 (25.58 - 25.94)
& 37.91 ± 0.00347 (37.66 - 38.16) / 15.66 ± 0.05578 (11.67 - 19.65) / 16.25 ± 0.00401 (15.97 - 16.54) \\

& Delay Randomization & 63.89 ± 0.00409 (63.59 - 64.18) / 74.07 ± 0.01673 (72.87 - 75.26) / 92.55 ± 0.00307 (92.33 - 92.77)
& 80.89 ± 0.00751 (80.35 - 81.43) / 83.1 ± 0.00553 (82.71 - 83.5) / 91.85 ± 0.00589 (91.43 - 92.27)
& 72.5 ± 0.00748 (71.96 - 73.03) / 75.11 ± 0.01717 (73.88 - 76.34) / 89.82 ± 0.00298 (89.6 - 90.03)
& 67.86 ± 0.00586 (67.44 - 68.28) / 74.74 ± 0.01732 (73.5 - 75.98) / 90.6 ± 0.00391 (90.32 - 90.88) \\
\hline

\end{tabular}
}
\captionsetup{justification=centering}
\caption*{Performance Metrics: Acc: Accuracy, Prec: Precision, Rec: Recall, F1: F1-Score, SD: Standard Deviation, 95\% CI: 95\% Confidence Interval. Datasets: d1 (IoT-AD), d2 (IoT Sentinel), d3 (UNSW). }
\end{table*}


\end{document}